\documentclass[twocolumn,aps,pra,superscriptaddress]{revtex4-2}
\usepackage[utf8]{inputenc}
\usepackage[T1]{fontenc}
\usepackage[english]{babel}
\usepackage{bm}
\usepackage[dvipsnames]{xcolor}
\usepackage{amsfonts}
\usepackage{ulem} 
\usepackage{tcolorbox}
\usepackage{wrapfig}
\usepackage{hyperref}
\usepackage{soul}
\usepackage{siunitx}
\usepackage{physics}
\AtBeginDocument{\RenewCommandCopy\qty\SI}
\usepackage{lineno}
\usepackage{tabularray}

\usepackage{lipsum}

\hypersetup{
colorlinks=true,
citecolor=blue,
linkcolor=red,
urlcolor=blue
 ,
 pdfmenubar=true
}

\usepackage[draft,todonotes={textsize=scriptsize},commandnameprefix=ifneeded]{changes}
\definechangesauthor[name={Cristian Bonato}, color=RoyalBlue]{CB}
\definechangesauthor[name={Pasquale Cilibrizzi}, color=cyan]{PC}

\definechangesauthor[name={Benedikt Tissot}, color=purple]{BT}
\definechangesauthor[name={Michael Trupke}, color=teal]{MT}

\setstcolor{blue}

\begin{document}

\title{Ultra-narrow inhomogeneous spectral distribution of telecom-wavelength vanadium centres in isotopically-enriched silicon carbide }

\author{Pasquale Cilibrizzi}
\affiliation{School of Engineering and Physical Sciences, SUPA, Heriot-Watt University, Edinburgh, EH14 4AS, United Kingdom}

\author{Muhammad Junaid Arshad}
\affiliation{School of Engineering and Physical Sciences, SUPA, Heriot-Watt University, Edinburgh, EH14 4AS, United Kingdom}

\author{Benedikt Tissot}
\affiliation{Department of Physics, University of Konstanz, D-78457 Konstanz, Germany}

\author{Nguyen Tien Son}
\affiliation{Department of Physics, Chemistry and Biology, Link\"oping University, SE-581 83 Link\"oping, Sweden}

\author{Ivan G. Ivanov}
\affiliation{Department of Physics, Chemistry and Biology, Link\"oping University, SE-581 83 Link\"oping, Sweden}

\author{Thomas Astner}
\affiliation{Institute for Quantum Optics and Quantum Information (IQOQI), Austrian Academy of Sciences, A-1090 Vienna, Austria}

\author{Philipp Koller}
\affiliation{Institute for Quantum Optics and Quantum Information (IQOQI), Austrian Academy of Sciences, A-1090 Vienna, Austria}

\author {Misagh Ghezellou}
\affiliation{Department of Physics, Chemistry and Biology, Link\"oping University, SE-581 83 Link\"oping, Sweden}

\author{Jawad Ul-Hassan}
\affiliation{Department of Physics, Chemistry and Biology, Link\"oping University, SE-581 83 Link\"oping, Sweden}

\author{Daniel White}
\affiliation{School of Engineering and Physical Sciences, SUPA, Heriot-Watt University, Edinburgh, EH14 4AS, United Kingdom}

\author{Christiaan Bekker}
\affiliation{School of Engineering and Physical Sciences, SUPA, Heriot-Watt University, Edinburgh, EH14 4AS, United Kingdom}

\author{Guido Burkard}
\affiliation{Department of Physics, University of Konstanz, D-78457 Konstanz, Germany}

\author{Michael Trupke}
\email{michael.trupke@oeaw.ac.at}
\affiliation{Institute for Quantum Optics and Quantum Information (IQOQI), Austrian Academy of Sciences, A-1090 Vienna, Austria}

\author{Cristian Bonato}
\email{c.bonato@hw.ac.uk}
\affiliation{School of Engineering and Physical Sciences, SUPA, Heriot-Watt University, Edinburgh, EH14 4AS, United Kingdom}

\begin{abstract}
 Spin-active quantum emitters have emerged as a leading platform for quantum technologies. However, one of their major limitations is the large spread in optical emission frequencies, which typically extends over tens of \unit{\GHz}. Here, we investigate single V\textsuperscript{4+} vanadium centres in 4H-SiC, which feature telecom-wavelength emission and a coherent $S=1/2$ spin state. We perform spectroscopy on single emitters and report the observation of spin-dependent optical transitions, a key requirement for spin-photon interfaces. By engineering the isotopic composition of the SiC matrix, we reduce the inhomogeneous spectral distribution of different emitters down to \num{100} \unit{\MHz}, significantly smaller than any other single quantum emitter.  Additionally, we tailor the dopant concentration to stabilise the telecom-wavelength V\textsuperscript{4+} charge state, thereby extending its lifetime by at least two orders of magnitude.  These results bolster the prospects for single V emitters in SiC as material nodes in scalable telecom quantum networks. 
 \end{abstract}

\maketitle

\section*{Introduction}

The demonstration of long-distance quantum networks represents a new paradigm for communication security \cite{Wehner_Science_Review_2018}. Single optically-active spin defects and impurities are strong candidates for physical nodes in such systems \cite{awschalom_quantum_2018, nemoto_quantum_networks_PRX, wolfowicz_quantum_2021}, and have been used in many of the most successful implementations of quantum networking primitives to date. A general architecture of quantum networks consists of an electronic spin associated with a point defect or impurity, interfaced to optical photons through spectrally stable spin-selective optical transitions. The electronic spin can couple to one or more nuclear spins \cite{taminiau_PRX_10spins, bourassa_entanglement_2020, murzakhanov_electronnuclear_2022}, which can serve as long-term quantum state storage sites and as ancillary qubits for implementing quantum error correction \cite{cramer_repeated_2016, muralidharan_repeaters_2016}. Examples of physical systems used to implement this architecture include the nitrogen-vacancy \cite{ humphreys_deterministic_2018, pompili_multinode_QN_2021}, silicon-vacancy \cite{SiV_Cavity_Lukin_2016, stas_Multi_QB_QN_2022} and tin vacancy \cite{parker2023diamond} centres in diamond; related systems in SiC \cite{fuchs_engineering_2015, Christle_SiC_2017, nagy_high-fidelity_2019, anderson_five-second_2022, babin_fabrication_2022}, silicon \cite{simmons_T_centre_PRXQ, Redjem_G_center_2020, hollenbach_wafer-scale_2022, dreau_G_centre} and 2D materials \cite{stern_room-temperature_2022}; and rare-earth ions in crystals \cite{ rare_earth_roadmap, raha_optical_2020, merkel_coherent_2020, afzelius_inhmg_broad_Yb_PRB2018}.

Quantum network applications can benefit from spin-photon interfaces in the telecom wavelength region, which are inherently compatible with standard optical fibre networks. While emission at alternative wavelengths can be converted to the telecom range by nonlinear optical processes \cite{Dreau_Hanson_2018, Weber_two-photon_2019, Rakonjac_ICFO_2021}, the additional hardware requirement reduces the overall system efficiency and can add noise. The spin-photon interface needs to be near lifetime-limited and spectrally stable to ensure high-visibility quantum interference. Additionally, to entangle different nodes in a quantum network, the inhomogeneous distribution of the optical transition frequencies for different emitters must be small to ensure indistinguishability of the emission, ideally close to the linewidth of individual defects. 

Quantum emitters typically exhibit a spread in emission frequencies induced by local differences in strain and electric field within the solid-state matrix. This problem has been addressed by tuning the optical transitions through the external application of either strain or electric fields \cite{Stark_Shift_2006,E_Tuning_Awschalom_2011}. Scaling up to multiple qubits is not trivial, however, due to the technical complexity of adding more and more electrical contacts or piezoelectric elements, while minimising cross-talk. An alternative solution is to exploit the same nonlinear process used to convert single photons to the telecom range, tuning the frequency of the pump laser to finely control the output telecom frequency so that all photons from all emitters are brought on resonance \cite{Weber_two-photon_2019}. This approach, however, adds substantial technological complexity for each emitter in the network.

Vanadium centres in silicon carbide (SiC) have recently received considerable interest for use in quantum networks, due to several attractive properties. Their emission directly in the O-band telecom range (\num{1260} \unit{\nm} to \num{1360} \unit{\nm}) \cite{spindlberger_optical_2019, wolfowicz_vanadium_2020} is directly compatible with standard fibres used in commercial optical telecommunications networks \cite{xiang2020tuneable} and allows interfacing with low-loss photonic circuitry \cite{wolfowicz_quantum_2021,fait_2021_finesse}. Direct emission in the telecom O-band removes the need for complex wavelength conversion and tuning hardware, and unlocks the possibility to wavelength-multiplex the quantum signal in the same standard telecom fibre with the classical signal in the C-band. Recent work has demonstrated long electron spin relaxation times (T$_1$) of up to \num{25} \unit{\s} \cite{astner_vanadium_2022},  ensemble dephasing times (T$_2^*$) of several microseconds and ensemble Hahn echo T$_2$ well in excess of 100 \unit{\micro \s} \cite{Hendriks_V_ensemble_2022} at cryogenic temperatures. 
SiC as a host material is widely used in high-power electronics, with established recipes for industrial-scale growth, doping, and fabrication, with very promising linear and nonlinear parameters for photonics \cite{castelletto_review_2022}. However, despite these encouraging features, several questions related to its suitability for application in a spin-photon interface remain open. In particular, the inhomogeneous spectral broadening in ensembles and single emitters is large compared to the natural linewidth of the defects, and stable emission of the neutral charge state in single defects has so far only been observed when applying a repump ultraviolet (UV) laser \cite{wolfowicz_vanadium_2020}.

Here, we investigate the optical and electronic properties of single neutral vanadium (V) centres in the 4H polytype of SiC (4H-SiC), providing experimental verification of spin-conserving optical transitions and their dependence on the applied magnetic field. We systematically study the optical emission of hundreds of V centres, comparing their distributions in standard and isotopically-enriched SiC. We find an ultra-narrow $\sim$\num{100} \unit{\MHz} inhomogeneous spectral distribution in isotopically-enriched SiC, compared to several GHz in SiC with a natural abundance of isotopes. This distribution is significantly smaller than any other single quantum emitter previously reported in the literature \cite{NV_centers_Strain_2019, Rogers_Inhomogenous_Si_in_Diamond_2014, SiV_Diamond_Inhomogeneous_2016,nagy_APL_2021,thomson_PRL_2018, thompson_SSRO_single_ions_science2020, simmons_single_T_NJP2021, Redjem_G_center_2020, dreau_G_centre}. We trace the origin of the ultra-narrow distribution to the reduction of local stress in the isotopically-enriched SiC, resulting from the nearly complete removal of any spread in the mass of elements in the crystal, i.e. the prevalence of different isotopes. Finally, we investigate the charge state dynamics of vanadium centres in SiC, demonstrating that, by tailoring the material purity, we can stabilise the required charge state for at least several seconds. Our results reveal the great potential of single V centres in SiC for scalable telecom quantum networking.

\section*{Results}
\subsection*{Spin-selective optical transitions}
A V centre forms when a V atom substitutes a silicon atom in the SiC crystal lattice \cite{Kunzer_V_Circula_Dichroism_6H_1993,spindlberger_optical_2019,wolfowicz_vanadium_2020}. In 4H-SiC, the neutral (V\textsuperscript{4+}) state is the only charge state that exhibits luminescence in the telecom region, featuring two zero-phonon lines (ZPL), denoted as $\alpha$ and $\beta$ \cite{spindlberger_optical_2019}, and an electronic spin $S=1/2$.

\begin{figure*}[!htbp]
\centering
\includegraphics[width=0.9\textwidth]{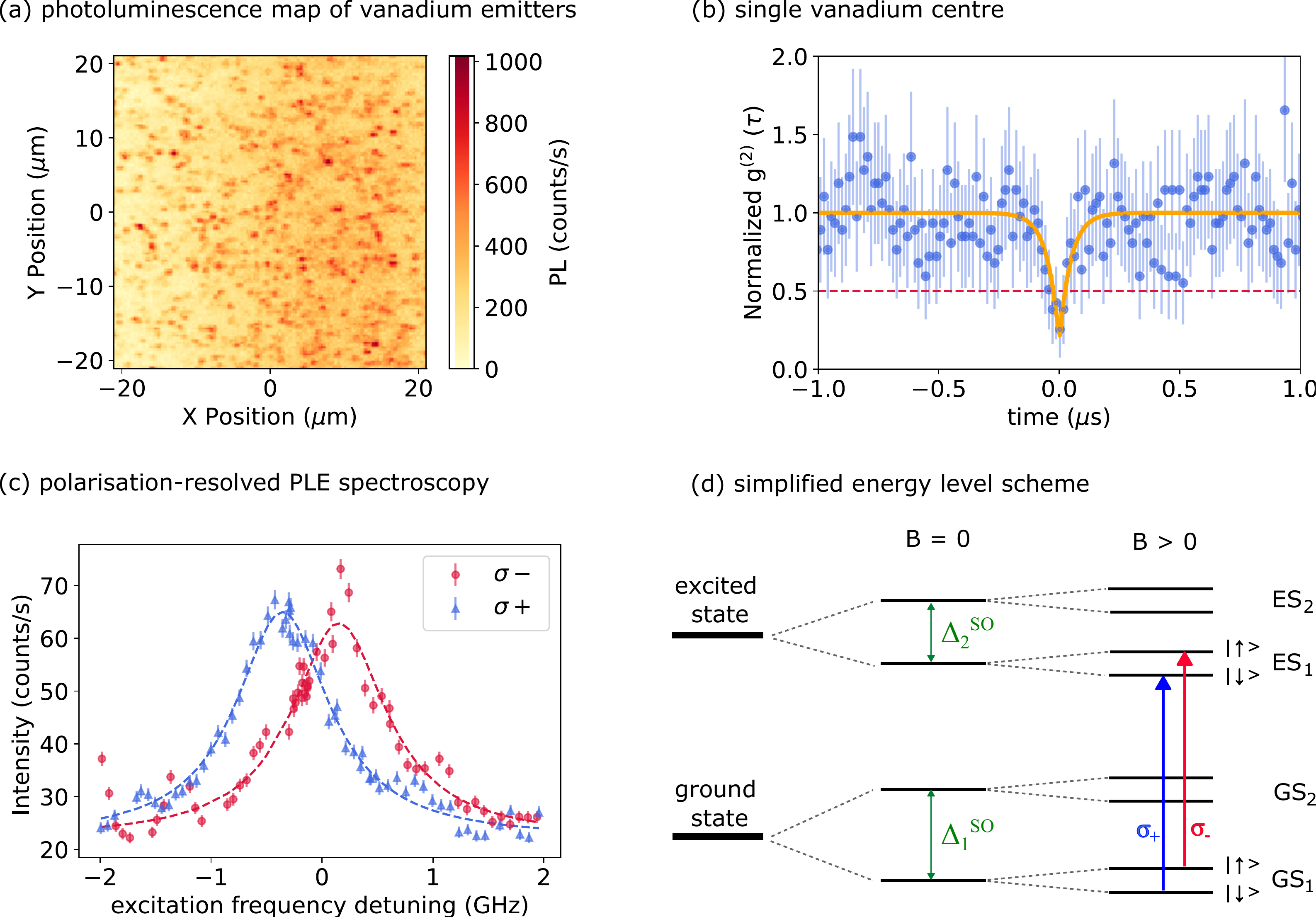}
\caption{\textbf{Spectroscopy of single V\textsuperscript{4+} centres in 4H-SiC.} 
\textbf{(a)} Photoluminescence (PL) intensity map acquired by scanning the telecom excitation laser (\SI{2.2}{\micro\watt}) across the sample on resonance with the $\alpha$ ZPL of the V centres (\num{1278.84} \unit{\nm}, \num{234.42457} \unit{\THz}). The V centres appear as localised PL spots. A \SI{14}{\micro\watt} repump is applied during the scan.
\textbf{(b)} Second-order autocorrelation measurement g$^{(2)}(\tau)$ for one of the spots, as a function of time delay ($\tau$). The data are not corrected for background counts. The measured $g^{(2)}$(0) = \num{0.255} shows that the spot corresponds to a single V centre. The yellow line corresponds to a single exponential fit function, f(x)=1 - A e$^{-\lvert t \rvert/t_1}$, with A = 0.828 $\pm$ 0.163 and t$_1$ = 0.048 $\pm$ 0.013 $\mu$s. The decay constant here is different than the known value of \num{167} \unit{\ns} for the emitter lifetime \cite{wolfowicz_vanadium_2020}, as it depends on the excitation optical power \cite{aharonovich_PRA_2010}. 
\textbf{(c)} Polarisation-resolved photoluminescence excitation (PLE) measurements, performed at an applied magnetic field of 1000 Gauss. For these measurements, we use excitation powers of \SI{14}{\micro\watt} for the repump laser and \SI{2.2}{\micro\watt} for the telecom laser. The integration time is set at 20 seconds per data point, with each point acquired three times and then averaged. The red circles and blue triangles correspond respectively to $\sigma_-$ and $\sigma_+$ circular polarisation. The frequency of the laser is expressed as detuning from the base frequency $f_0$ = \num{234425} \unit{\GHz}. \textbf{(d)} Simplified energy level diagram for V centres in SiC. The degeneracy of the ground (GS) and excited (ES) states, set by the C$_{3v}$ symmetry, are broken by spin-orbit coupling ($\Delta_1^{SO}$ and $\Delta_2^{SO}$)  \cite{tissot-electronic-2021,tissot_nuclear_2022}. An applied magnetic field $B$ further removes the degeneracy between the electron spin levels.  The blue and red arrows indicate the excitation of the two spin-dependent transitions with circularly polarised light ($\sigma_+$ and $\sigma_-$, respectively). The error bars in (b) and (c) correspond to the Poisson noise on the photon counts.}
\label{fig:1}
\end{figure*}

We identify V\textsuperscript{4+} centres by confocal spectroscopy at \num{4.3} \unit{\K} (see Methods and Supplementary Note 1), using a narrowband tuneable CW laser (1278.8 nm) to resonantly excite the $\alpha$ zero-phonon line (ZPL) of V\textsuperscript{4+} \cite{spindlberger_optical_2019}, while detecting the phonon sideband emission (\num{1300} \unit{\nm} - \num{1600} \unit{\nm}). The typical excitation powers range from \SI{1}{\micro\watt} to \SI{4}{\micro\watt} depending on the specific experiments. We use a green repump laser (\num{520} \unit{\nm}, \SI{14}{\micro\watt}) to compensate for laser-induced ionisation \cite{wolfowicz_vanadium_2020}. The necessity of a repump laser is discussed in more detail in the section concerning the stability of the neutral charge state and in Supplementary Note 2. By scanning the excitation lasers across an isotopically-enriched sample (sample A, described in Methods), we obtain bright photoluminescence (PL) spots as shown in Fig. \ref{fig:1} (a). An automated detection algorithm (see Methods) finds 389 spots in Fig. \ref{fig:1} (a), corresponding to \num{0.29} spots/\unit{\micro\meter\squared} over a map area of \num{1600} \unit{\micro\meter\squared}. We confirm the single emitter nature of one of the spots by g$^{(2)}(\tau)$ auto-correlation measurements, observing a g$^{(2)}(0)$ = 0.255 $\pm$ 0.180 at zero delay (Fig. \ref{fig:1} (b)). 

We investigate the single V\textsuperscript{4+} centre through photoluminescence excitation (PLE) spectroscopy, by scanning the telecom laser frequency across the $\alpha$ ZPL with selected $\sigma_+$ ($\sigma_-$) circular polarisation at a magnetic field of \num{1000} Gauss, oriented along the SiC c-axis (see Methods and SI for details). The PLE measurements reveal two distinct peaks with orthogonal circular polarisations (Fig. \ref{fig:1} (c)). We fit the PLE spectra with Lorentzian functions, centred at \num{0.155} $\pm$ \num{0.015} \unit{\GHz} ($\sigma_+$) and \num{-0.335} $\pm$ \num{0.011} \unit{\GHz} ($\sigma_-$). The relative Zeeman splitting of \num{510} \unit{\MHz}, extracted from the fittings, is in agreement with theoretical predictions \cite{tissot-electronic-2021} (see Supplementary Note 3 for details). The $\sigma_+$ and $\sigma_-$ peaks in Fig. \ref{fig:1} (c), result from circular polarisation-dependent selection rules \cite{tissot_nuclear_2022} and are associated with the spin-conserving transitions between the first ground (GS$_1$) and first excited state (ES$_1$), as schematically shown in Fig. \ref{fig:1} (d). The electronic structure of V\textsuperscript{4+} is determined by a single active electron localised in a $d$ orbital, which possesses C$_{3v}$ symmetry imposed by the SiC crystal field. The orbital degeneracy is lifted by spin-orbit coupling into Kramers doublets (KDs): pairs of degenerate states connected through time inversion. The degeneracy of the Kramers doublets can be lifted by an applied magnetic field \cite{gilardoni_hyperfine_mediated_2021,tissot-electronic-2021, tissot_nuclear_2022} (see Supplementary Note 3 for details). As a consequence, spin-selective optical transitions between KDs in the ground and excited states addressable by circularly-polarised excitation light \cite{tissot-electronic-2021}  are expected.

\begin{figure}[!htbp]
\includegraphics[width=8.5cm]{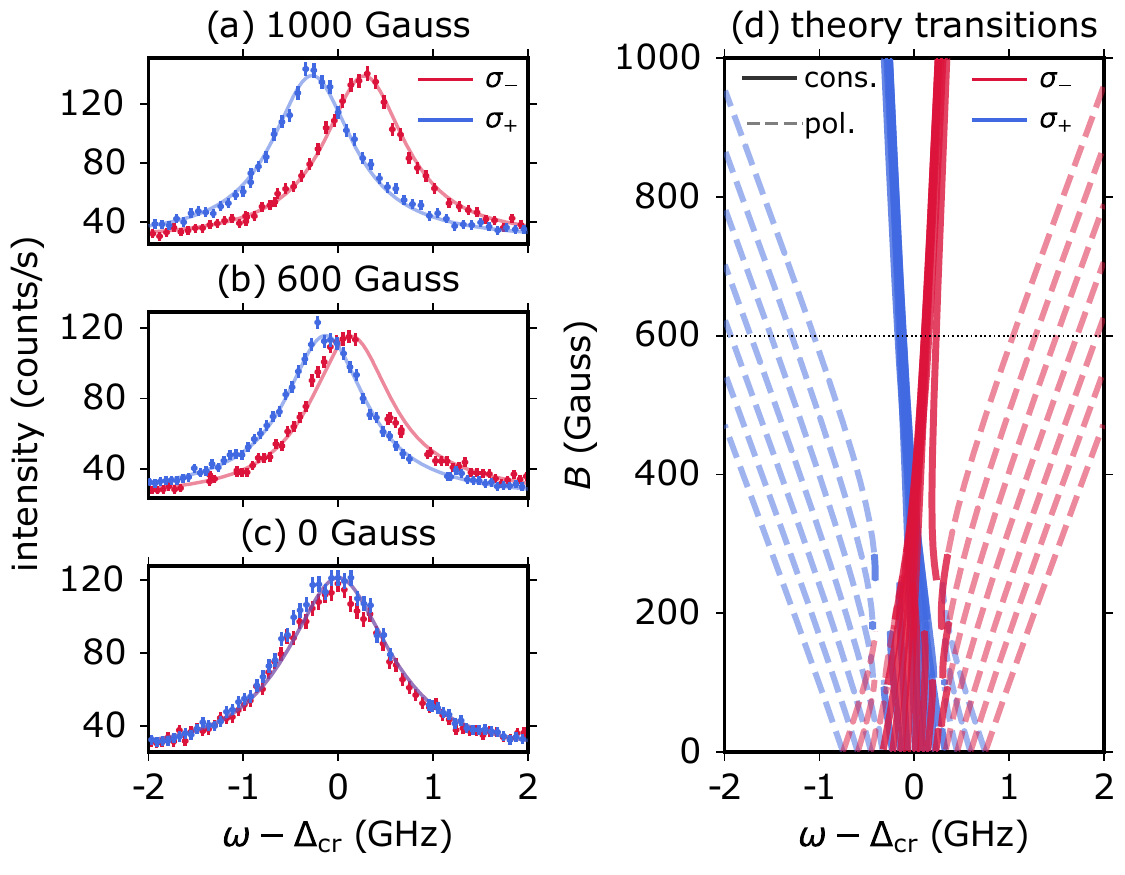}
\caption{\textbf{PLE spectroscopy as a function of the applied magnetic field.} \textbf{(a)},\textbf{(b)},\textbf{(c)} PLE spectroscopy of the same PL spot, with selected $\sigma_-$ (red) and $\sigma_+$ (blue) circular polarisation excitation, under different magnetic fields applied along the sample c-axis (\num{0} Gauss, \num{600} Gauss and \num{1000} Gauss, respectively). For this experiment, we do not confirm the single nature of the emitters with g$^{(2)}(\tau)$ measurements. Here we use \SI{14}{\micro\watt} for the repump laser and \SI{1}{\micro\watt} for the telecom laser. The integration time is 15 seconds per data point, and each point is acquired three times and subsequently averaged. The error bars correspond to the Poisson noise of the photon counts. \textbf{(d)} Theoretical spin-conserving (cons.) and nuclear spin-polarising (pol.) hyperfine allowed transitions predicted by the model outlined in Supplementary Note 3 and using the hyperfine parameters as in \cite{wolfowicz_vanadium_2020, tissot-hyperfine-2021, astner_vanadium_2022}. The electron-spin-conserving transitions between the ES$_1$ and GS$_1$ (Fig. \ref{fig:1} (d)) electron levels (solid lines) narrow for higher magnetic fields and conserve the nuclear and Kramsers doublet (KD) spin, while the hyperfine allowed transitions (dashed lines) flip the KD spin and can polarise the nuclear spin. The fit (solid lines in (a)-(c)) allows for different amplitudes and backgrounds for each of the measurements, but shares the same width for all transition frequencies (see (d)) \num{1038} $\pm$ \num{7} \unit{\MHz} and central transition frequency $\Delta_{cr}$ = \num{234425594} $\pm$ \num{4} \unit{\MHz}. The polarisation information is encoded in the curve colour (see legend).}
\label{fig:PLE_vs_B}
\end{figure}

We further investigate the magnetic field dependence of the optical transitions through polarisation-resolved ($\sigma_+$ and $\sigma_-$) PLE spectroscopy as a function of the magnetic field strength $B$ applied along the c-axis (Fig. \ref{fig:PLE_vs_B}). \\
In the absence of an external magnetic field (B = \num{0} Gauss), the two $\sigma_+$ and $\sigma_-$ circularly polarised PLE spectra are superposed with no visible splitting. At B = \num{600} Gauss, a splitting of \num{220} \unit{\MHz} becomes visible, which increases to \num{510} \unit{\MHz} at B = \num{1000} Gauss. Remarkably, the linewidth of the resonance decreases for a higher applied magnetic field.
We compare our data with the theoretical models (Fig.~\ref{fig:PLE_vs_B} (d)) based on the $d$-orbital configuration and $C_{3v}$-symmetry of the centre \cite{tissot-electronic-2021,tissot-hyperfine-2021,tissot_nuclear_2022}, calculating the relative frequencies and strengths of the hyperfine transitions using previously measured experimental values \cite{wolfowicz_vanadium_2020,tissot-hyperfine-2021,astner_vanadium_2022} (adapted to the symmetry-based hyperfine model as in \cite{tissot_nuclear_2022}, see Supplementary Note 3). 
We assume a Lorentzian shape with the same full-width $\gamma$ for all hyperfine allowed transitions. The width $\gamma \approx$ 
\num{1} \unit{\GHz}, the defect-dependent central (or zero-field) transition frequency $\Delta_{cr}$, as well as parameters describing the background offset and the amplitude of the peak are fit parameters.
Further details on the fit and model are presented in Supplementary Note 3.

In simple terms, the decrease in linewidth with increasing magnetic field can be understood from the fact that the off-diagonal coupling, which is of a vastly different form for the involved KDs, is suppressed under large magnetic fields and, if the $zz$ components share the same sign, the diagonal coupling leads to a similar hyperfine splitting.
Therefore, the model (using the literature values for the hyperfine parameters of ''bulk'' defects much deeper than the size of their wavefunctions \cite{csore-2020-ab_initio}, with intact symmetry) predicts that the electron-spin-conserving transitions will converge at higher magnetic fields, resulting in a narrowing of the linewidth.
The splitting of the two peaks can be approximated by $\mu_B |g_e-g_g| B$ for large magnetic fields $B \gg 100$ Gauss, in agreement with our experimental measurement.

\subsection*{Ultra-narrow inhomogeneous spectral distribution}

We next investigate the spectral inhomogeneous distribution of the ZPLs corresponding to different V\textsuperscript{4+} centres. Previous experiments on V ensembles \cite{wolfowicz_vanadium_2020} 
have highlighted an asymmetry in the ZPL, with a longer tail and duplicated lines at higher frequencies. This has been attributed to the isotope shift from nearby $^{29}$Si, $^{30}$Si and $^{13}$C isotopes. The isotope shift is a change in frequency of the ZPL given by a variation in the mass of nearby atoms, corresponding to heavier or lighter isotopes, that changes the local strain or bandgap \cite{Isotope_shift_Theory_1975, lawson_h2_1992, RevModPhys_Isotope_shift_2005}.
Linewidth fitting of ensemble optical spectroscopy data \cite{wolfowicz_vanadium_2020} has given a shift of \num{22} $\pm$ \num{3} \unit{\GHz} per unit mass for nearest-neighbour carbon isotopes and \num{2.0} $\pm$ \num{0.5} \unit{\GHz} per unit mass for silicon isotopes. 
Further evidence of this was recently provided by optically-detected magnetic resonance (ODMR) measurements \cite{Hendriks_V_ensemble_2022}, showing a change in the ODMR spectrum as the optical excitation wavelength of a V ensemble was tuned.
In particular, while excitation at a wavelength of \num{1278.86} \unit{\nm} (a detuning of about \num{-3} \unit{\GHz} compared to our base frequency $f_0$) only showed a dip related to the V centre electron spin resonance, excitation at \num{1278.76} \unit{\nm}
(about +\num{15} \unit{GHz} from $f_0$) evidenced the presence of side peaks consistent with hyperfine coupling to neighbouring $^{29}$Si atomic nuclei.
This suggests that when SiC is isotopically enriched, removing most $^{29}$Si, $^{30}$Si and $^{13}$C isotopes, the inhomogeneous distribution of the ZPL should become narrower compared to the case of isotope composition according to natural abundance. 

We investigate this quantitatively by performing a sequence of PLE maps, each at different excitation frequency, in two different SiC samples, one isotopically enriched (sample A), and one featuring a natural abundance of Si and C isotopes (sample B). Both samples are described in detail in the Methods. A sub-set of the maps is shown in Fig. \ref{fig:inhomogeneous_broadening} (a), with the full sequence reported in the Supplementary Note 4 B.
To quantify the spectral distribution of the central frequency of the V emission, we perform a statistical analysis of the PLE spots in the maps. We process the series of maps by automatically detecting spots of sizes compatible with the diffraction limit, and fitting the PLE intensity for each spot, as a function of excitation frequency, with a Gaussian function to extract the centre frequency (details in the Supplementary Note 4). Histograms for the inhomogeneous distributions of central frequencies in the two samples are shown in Fig. \ref{fig:inhomogeneous_broadening}(c). In sample B, with a natural abundance of isotopes, the frequencies span across several GHz, with a few peaks in the distribution, compatible with the $\sim 2$ GHz per unit mass expected for Si isotopes \cite{wolfowicz_vanadium_2020}. In future work, it will be interesting to correlate optical spectroscopy with optically-detected electron spin resonance measurements to investigate the hyperfine coupling for each different detuning of the ZPL. This was not possible here, since we could not detect any optically-detected magnetic resonance on single centres, likely due to the short electron spin relaxation timescales at \num{4} \unit{\K} for V \cite{astner_vanadium_2022}. In the isotopically-enriched sample, the central frequencies exhibit a much narrower distribution, with standard deviation $\sim$\num{100} \unit{\MHz}. Over three regions in sample A, the central frequencies of the distributions are all quite close: $f_A$ = \num{227} $\pm$ \num{105} \unit{\MHz}, $f_B$ = \num{-41} $\pm$ \num{109} \unit{\MHz}, $f_C$ = \num{-25} $\pm$ \num{76} \unit{\MHz}, comparable to the absolute accuracy of the wavemeter ($\sim$ 150 MHz). A fourth measurement in a different region of the crystal shows a detuning of \num{700} \unit{\MHz}. This measurement was taken after a thermal cycle of warm-up and cool-down, which may have affected the sample strain. However, within each region, we observe the same \num{100} \unit{\MHz} ultra-narrow spectral distribution.

We can trace the origin of the frequency shifts in the optical transitions of V\textsuperscript{4+} centres to the local strain produced by nearby isotopes \cite{wolfowicz_quantum_2021}. 
Theoretically, we will treat the coupling to strain analogously to the coupling to electric fields (see \cite{tissot-electronic-2021}). This
is possible for several reasons: the symmetry arguments were used on the orbital level, the fact that the strain coupling fulfils time-reversal symmetry and the possibility to assign (combined) strain tensor elements to the same irreducible representations as the electric field components \cite{udvarhelyi-NV_strain-2018}. In particular, there are two strain tensor elements (i.e., \(\epsilon_{zz}\) and \(\epsilon_{xx}+\epsilon_{yy}\)) that transform like an electric field in the \(z\)-direction within the $C_{3v}$ symmetry which can directly influence the energy spacing of the KDs \cite{tissot_strain_2023}.

A local inhomogeneous distribution on the order of \num{100} \unit{\MHz} compares very favourably with other quantum emitters. This is several orders of magnitudes smaller than literature values for single nitrogen-vacancy (NV) centres in diamond ($\sim$ \num{40}-\num{50} \unit{\GHz} \cite{NV_centers_Strain_2019}). A narrower inhomogeneous distribution, on the order of \num{10}-\num{20} \unit{\GHz}, has been reported for a single silicon-vacancy centre (SiV) in diamond \cite{Rogers_Inhomogenous_Si_in_Diamond_2014} and silicon carbide (V\textsubscript{Si}) \cite{nagy_APL_2021}. V in SiC compares favourably even to single rare-earth ions and single emitters in silicon, which can feature narrow distributions down to the MHz-GHz level \cite{sellars_inhmg_broad_Eu_PRL2016, afzelius_inhmg_broad_Yb_PRB2018,thompson_SSRO_single_ions_science2020, serrano_ultra-narrow_2022}. A more informative figure of merit could be the ratio $\eta$ between inhomogeneous broadening and the transform-limited linewidth, which quantifies by how many linewidths the frequencies of two emitters need to be shifted to be brought into resonance. According to this metric (Supplementary Table 3), V in SiC performs on par with the SiV centre in diamond ($\eta \sim$ 100), a system protected against electric fields and strain by inversion symmetry, more than an order of magnitude better than the NV centre in diamond ($\eta \sim$ 3400). 

 \begin{figure*}[!htbp] 
 \centering
 \includegraphics[width=0.95 \textwidth]{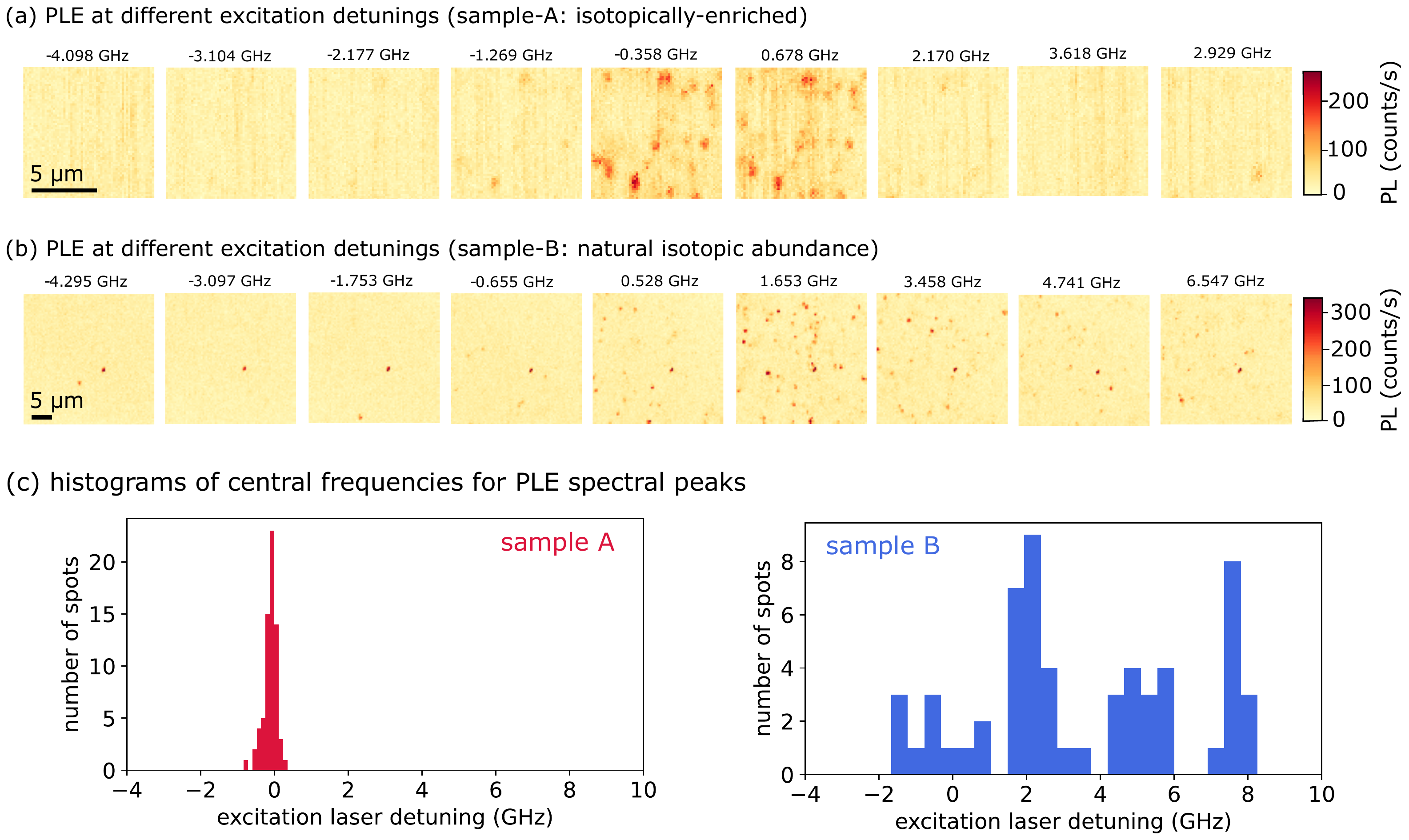}
 \caption{ \textbf{Inhomogeneous spectral distribution of the V centres in SiC with different isotopical composition.} \textbf{(a)}, \textbf{(b)} Sequence of PLE maps at different detunings of the excitation laser (\SI{14}{\micro\watt} for the repump laser,  \SI{4}{\micro\watt}  for the telecom laser), respectively for the isotopically-enriched sample (A) and the natural-abundance sample (B). In the isotopically-enriched sample the vast majority of emitters appear only in a narrow frequency range. Maps in (a) are 10 $\mu$m $\times$ 10 $\mu$m, maps in (b) 30 $\mu$m $\times$ 30 $\mu$m. We acquired larger maps for sample B because the concentration of spots in each map is smaller than that for sample A (due to the larger frequency spread). \textbf{(c)} Histograms for the central frequencies of 181 PL spots associated to V centres in the natural-abundance (right) and 61 spots in isotopically-enriched (left) samples. In the sample with a natural abundance of Si and C isotopes, the V\textsuperscript{4+} centres are spectrally spread over several GHz, presenting a distribution with multiple peaks. In contrast, the distribution is much narrower in the isotopically-enriched sample, with a standard deviation of about 100 MHz.
}
\label{fig:inhomogeneous_broadening}
\end{figure*}

\subsection*{Stability of the neutral charge state}
\label{Charge_State}

Finally, we investigate the stability of the neutral V\textsuperscript{4+} charge state. In 4H-SiC, V can exist in three possible charge states within the bandgap: V\textsuperscript{3+} (negatively charged), V\textsuperscript{4+} (neutral) and V\textsuperscript{5+} (positively charged) \cite{baur_transition_1997}, with only the V\textsuperscript{4+} state featuring optical emission in the telecom region and a $S=1/2$ electronic spin. The lifetime of the V\textsuperscript{4+} charge state is important as it sets a limit for the use of V centres in quantum technology applications.

The stabilisation of a given charge state for a deep-level defect depends on the local Fermi level and the complex interplay with other nearby dopants/defects that act as electron donors or acceptors. The charge state transition levels for V and the other main defects in SiC, such as the carbon vacancy (V\textsubscript{C}) and the divacancy (V\textsubscript{Si}V\textsubscript{C}), are displayed in Fig. \,\ref{fig:3} (a). We include a discussion in Supplementary Note 2 about how these levels have been determined in the literature. 

\begin{figure*}[htpb] 
\centering
\includegraphics[width=1 \textwidth]{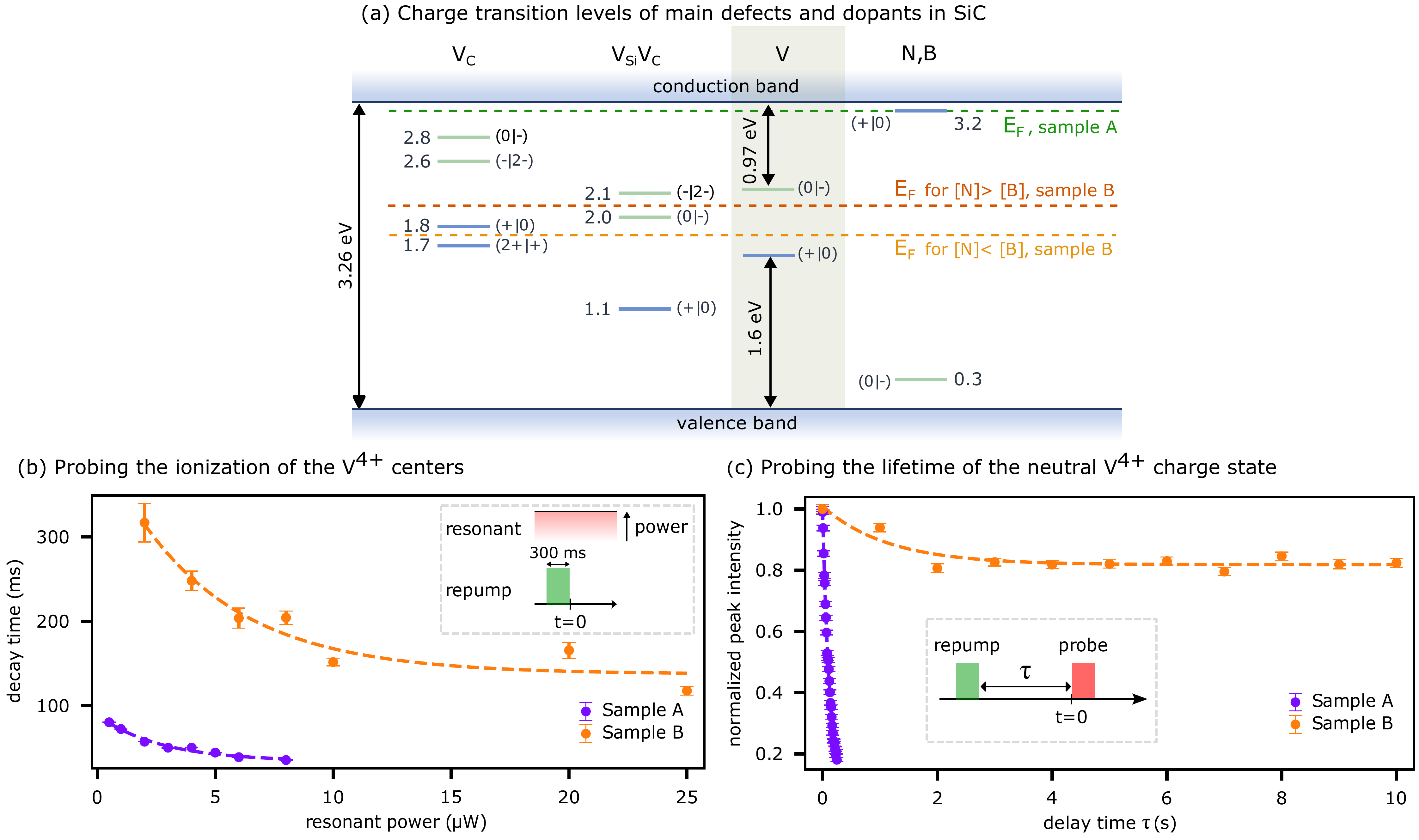}
\caption{\textbf{Stability of the V\textsuperscript{4+} charge state.} \textbf{(a)} Energy diagram showing the charge transition levels for V centres, the main dopants nitrogen (N) and boron (B), and intrinsic defects, such as carbon vacancy (V\textsubscript{C}) and divacancy, (V\textsubscript{Si}V\textsubscript{C}) involved in the V charge state dynamics in SiC. A transition between the positively-charged and neutral states, for example, is labelled as (+/0). V\textsuperscript{4+} corresponds to the neutral (0) and V\textsuperscript{3+} to the negative (-) charge state. In sample A (n-type), the N shallow donor can compensate the acceptor levels of shallow B and V\textsubscript{C}, pinning the Fermi level (E$_F$) at the N shallow donor level and stabilising V\textsuperscript{3+}. In sample B (semi-insulating), the E$_F$ is located at the $\left(0|-\right)$ acceptor level of the divacancy if the concentration of the N donor is larger than that of the B acceptor or, vice versa, at the $\left(+|0\right)$ donor level of V\textsubscript{C}. In both cases, V\textsuperscript{4+} is stabilised at equilibrium.  \textbf{(b)} Decay of the V\textsuperscript{4+} charge state at increasing powers of the resonant CW excitation. V\textsuperscript{4+} is prepared by a \num{300} \unit{\ms} repump pulse and the PL decay over time is recorded as a function of the resonant laser power (see pulse sequence inset). The ionisation induced by the resonant excitation follows an exponential decay of the count rate for both samples (see Supplementary Note 2), with a weaker decay in sample B, where the Fermi level stabilises V\textsuperscript{4+}. \textbf{(c)} Lifetime of the V\textsuperscript{4+} charge state when no optical excitation is present. The system is prepared in the V\textsuperscript{4+} state by a repump pulse and the PL emission intensity is then recorded as a function of the delay ($\tau$) between the repump and the probe. The PL peak intensity, corresponding to the occupation of the V\textsuperscript{4+} charge state, is plotted against the delay $\tau$ for both samples (see Supplementary Note 2). In sample A, the V\textsuperscript{4+} charge state decays rapidly with a timescale of \num{129} $\pm$ \num{6} \unit{\ms}, while it survives for several seconds in sample B. Errors bars in (a) and (b) correspond to the standard deviations extracted from the fits.}
\label{fig:3}
\end{figure*}

As seen in previous experiments \cite{wolfowicz_vanadium_2020, astner_vanadium_2022}, resonant excitation of the $\sim$\num{1278} \unit{\nm} line resulted in luminescence quenching due to the change in the charge state of V\textsuperscript{4+}, which is restored by a non-resonant (UV or green) repump laser.  Here we characterise the charge dynamics of the V\textsuperscript{4+} state though two experiments, each performed on both samples A and B, which feature two different nitrogen and boron doping concentrations (described in detail in the Methods). For sample A, we expect the Fermi level to be pinned at the N shallow donor level, while for sample B (semi-insulating) it is situated approximately in the middle of the bandgap (as shown in Fig \ref{fig:3}(a)).

In the first experiment (Fig. \ref{fig:3}(b)), we probe the ionisation of the V\textsuperscript{4+} centre due to the telecom (resonant) excitation laser, with the scheme shown  in the inset of Fig. \ref{fig:3}(b). We prepare the V\textsuperscript{4+} state with a green laser pulse (\num{300} \unit{\ms}), and detect the PL decay at different CW telecom laser powers. We fit the decays at different excitation powers with a single exponential, and plot the decay rates, extracted from the previous fits (shown in Supplementary Note 2), as a function of the resonant excitation power (Fig. \,\ref{fig:3} (b)). In both samples, the resonant excitation laser induces quenching of PL from V\textsuperscript{4+}, but the effect is more pronounced in sample A.

In the second experiment, we assess the lifetime of the neutral V\textsuperscript{4+} charge state when no optical excitation is present (Fig. \,\ref{fig:3} (c)). This is relevant, for example, for quantum memory experiments, when optical excitation is only used for short spin initialisation and readout, and the spin preserves a quantum state in the dark.
To determine this timescale, we initialise the V\textsuperscript{4+} state with a green laser pulse (\num{300} \unit{\ms}) and subsequently apply a readout pulse with the resonant laser (probe, \num{300} \unit{\ms}), after a delay $\tau$ (see inset in (Fig. \,\ref{fig:3} (c)). By increasing the delay $\tau$ and monitoring the change in PL during the readout pulse, we measure the PL decay at different delays. We then fit the decays at different delays ($\tau$) using a single exponential, and plot the peak intensities extracted from the fits as a function of the delay time $\tau$ (Fig. \,\ref{fig:3} (c)).
For sample B in Fig. \,\ref{fig:3} (c), we repeated the experiment twice, and plotted the average of the two measurements. We observe that, even in the absence of laser excitation, the PL peak intensity decays exponentially in sample A, with a decay constant of \num{129} $\pm$ \num{6} \unit{\ms} (violet curve). For sample B, this timescale is considerably extended, up to several seconds (orange curve). Note that this is a lower bound as, at these timescales, the decay is likely determined by unwanted excitation by the telecom laser leaking through the acousto-optic modulator when turned off, due to a non-ideal extinction ratio.

These experimental observations can be understood as being due to the different Fermi level in sample B compared to sample A, which improves the stability of the V$^{4+}$ charge state in sample B. To enable discrimination of single vanadium emitters, we used an implantation dose of \num{1}$\times$\num{10}$^8$ \unit{cm.^{-2}} in both samples (see Methods). At this low dose, the Fermi level is not affected by V dopants but rather by the concentration of residual nitrogen (N) and boron (B), and other intrinsic defects such as divacancies and carbon vacancies (see Methods for details). Specifically, in sample A, the Fermi level is pinned to the donor level of N (see  Fig. \,\ref{fig:3} (a) "E$_F$, sample A"). In such n-type sample, V is expected to exist in the negative charge state V\textsuperscript{3+} at equilibrium. 

The re-pump laser induces excitation of electrons from the acceptor levels of V (0|-) to the conduction band, thus enabling charge state conversion from V\textsuperscript{3+} (dark) to V\textsuperscript{4+} (bright). The decay that we observe in Fig. \ref{fig:3} (c) can be explained as an effect of the resonant laser used to probe V\textsuperscript{4+}, leading to transformation of V\textsuperscript{4+} into V\textsuperscript{3+}. Indeed, the telecom resonant excitation pumps electrons from the shallow N donor and V\textsubscript{C} acceptor levels into the conduction band, which can be re-captured by V\textsuperscript{4+} to create V\textsuperscript{3+}, leading to the measured exponential decays of its PL on a 129 ms timescale. We notice that the resonant excitation alone is most likely insufficient to reactivate V\textsuperscript{4+}, which stipulates the need of repump laser of higher energy (see Supplementary Note 2B).

Furthermore, the higher the resonant laser power, the higher the excitation rate of free electrons, leading to higher free-electron concentration and increased capture rate to the V\textsuperscript{4+} centres. This explains  the observed decrease in the decay time observed in Fig. 4(b).

In sample B, the residual N donors and B acceptors have comparable concentrations, in the low $10^{15}$ \unit{\per\cubic\cm} range, and the carbon vacancy and divacancy are the dominant intrinsic defects (with a concentration in the high $10^{15}$ \unit{\per\cubic\cm} range).  The insulating character of this sample sets the Fermi level to the middle of the gap, stabilising V\textsuperscript{4+}. A more detailed discussion about the expected charge state dynamics at the microscopic level can be found in Supplementary Note 2.

\section*{Discussion}

In this work, we have investigated several open questions related to the spin-active V centre in SiC. We have utilised two types of SiC, one with isotopically-enriched composition but slightly n-doped, and one high-purity, semi-insulating sample with natural isotopic composition. We have provided experimental verification of spin-dependent optical transitions in V, investigating their dependence on the applied magnetic field and confirming recent theoretical predictions on its electronic structure \cite{tissot-electronic-2021}. Remarkably, we observe an ultra-narrow $\sim$\num{100} \unit{\MHz} spectral inhomogeneous distribution in the isotopically-enriched sample: the comparison in the Supplementary Table 3 highlights that this is the narrowest observed for any single solid-state emitter. This is achieved through isotopic enrichment of the SiC host which removes the isotope shift effect, with different isotopes locally inducing a different frequency shift in the emission \cite{wolfowicz_vanadium_2020}. A narrow inhomogeneous distribution is crucial to create remote entanglement between different emitters in the nodes of a quantum network, as high-visibility quantum interference requires perfect indistinguishability between the photons. It could also be beneficial for super-radiance and collective emission \cite{koong_coherence_2022, lukin_two-emitter_2023} experiments, that similarly require indistinguishability. This property could also be used to pre-select emitters with a specific configuration of nearby nuclear spins by optical spectroscopy, facilitating the implementation of specific emitters in a quantum memory.  
Finally, we engineer the material doping level to stabilise the V\textsuperscript{4+} charge state, which features telecom emission and a spin $S=1/2$ state. Doping level and isotopic composition are independent parameters, that can be combined to simultaneously achieve stabilisation of the correct charge state and narrow inhomogeneous distribution on the same sample.

One open question is related to the linewidth of the observed optical transitions. We measure a linewidth of about \num{600} \unit{\MHz} at the lowest excitation power, in contrast to a lifetime limit of less than \num{1} \unit{\MHz}. The electronic optical transition is broadened with respect to the lifetime limit by the presence of 16 hyperfine levels corresponding to the interaction with the $I = 7/2$ $^{51}$V nuclear spin, with a hyperfine interaction on the order of \num{200} \unit{\MHz} (see Supplementary Table 3). Including the hyperfine transitions, however, should result in a set of multiple separated MHz-linewidth lines, that would enable direct optical access to the nuclear spin. Temperature can also affect the linewidth, through phonon-assisted dephasing. In our setup, experiments are limited to about \num{4} \unit{\K} and additional measurements at lower temperature are required to fully assess the impact of phonon dephasing.
A further mechanism could be related to spectral diffusion induced by fluctuating electric fields associated with other defects in the vicinity of the V centres. The implantation process can result in lattice damage, especially for heavier atomic species like V. While we have annealed both samples at a very high temperature after implantation (see Methods), this may have not fully repaired the SiC crystal lattice. A possible way to assess and address the contribution of this factor may be to embed the vanadium centres in a p-i-n diode structure and apply a large electric field to empty all of the charge traps, stabilising the local electric field environment \cite{anderson_electrical_2019}.

Single V centres can be detected without any photonic structure, in contrast for example to most rare-earth ion emitters. Through resonant excitation, we observe count rates on the order of 150 counts per second in the phonon sideband. With a Debye-Waller factor of about 0.3 \cite{spindlberger_optical_2019}, this corresponds to about 75 counts per second in the ZPL. The low count rate is not completely explained by the relatively long excited-state lifetime (\num{167} \unit{\ns} \cite{wolfowicz_vanadium_2020}), suggesting a reduced quantum efficiency of the emission. Given the rich hyperfine structure of $I=7/2$ \textsuperscript{51}V nuclear spins, resonant optical excitation could lead to pumping into a dark state. If this is the case, nuclear spin polarisation \cite{tissot_nuclear_2022} would boost counts and additionally provide access to a fast quantum memory, with hyperfine interaction on the order of \num{200} \unit{\MHz}. Optical collection could be enhanced through the use of photonic structures such as solid immersion lenses \cite{Tihan_SIL_2023} or nanopillars \cite{castelletto_review_2022}. Resonant photonic structures could further enhance emission: recent work on rare-earth ions has shown that Purcell factors exceeding few hundreds can be achieved both by photonic crystals \cite{raha_optical_2020}, open microcavities \cite{merkel_coherent_2020} and plasmonic waveguides \cite{gusken_emission_2023}.  If the spin-photon interface is implemented exploiting the circular polarisation of the optical transition, it is important to preserve the polarisation degeneracy of the cavity modes, which requires controlling shape and birefringence in the case of micropillars or open cavities \cite{gudat_permanent_tuning_2011}, or utilising symmetric geometries in the case of photonic crystals \cite{hagemeier_h1_2012}.

In conclusion, our results, in combination with recent measurements of long spin relaxation and coherence lifetimes below T=\num{2} \unit{\K} \cite{Hendriks_V_ensemble_2022}, reveal the potential of single V centres in SiC for quantum networking and as sensitive probes of their crystalline environment. By engineering the isotopic composition and purity of the SiC host material, we have shown that it is possible to stabilise the $S=1/2$ neutral charge state and narrow the spectral distribution of different emitters, facilitating the realisation of entanglement between multiple V centres in telecom-wavelength quantum networks.

\section*{Methods}
\textbf{Samples}. In this work we utilise two SiC samples, labelled as "A" and "B". 

Sample A is a $\sim$ \num{110} \unit{\mu\meter} thick isotopically-enriched 4H-\textsuperscript{28}Si\textsuperscript{12}C layer, grown by chemical vapour deposition (CVD) on the Si-face of a standard 4-degrees off-axis (0001) 4H-SiC substrate. The isotope purity is estimated to be ~ 99.85\% for \textsuperscript{28}Si and ~ 99.98\% for \textsuperscript{12}C, which was confirmed by secondary ion mass spectroscopy (SIMS) for one of the wafers in the series. The current-voltage measurements using a mercury probe station shows that the layer is n-type with a free carrier concentration of $\sim$ \num{6}$\times$\num{10}$^{13}$ \unit{\per\cubic\cm}, which is close to the concentration of the residual N shallow donor of $\sim$ \num{3.5}$\times$\num{10}$^{13}$ \unit{\per\cubic\cm} as determined from low-temperature photoluminescence (PL). Due to contamination from the susceptor, the concentration of the B shallow acceptor is expected to be in the low \num{10}$^{13}$ \unit{\per\cubic\cm} range. Deep level transient spectroscopy (DLTS) measurements show that the dominant electron trap in the layer is related to the carbon vacancy V\textsubscript{C} with a concentration in the range of low \num{10}$^{13}$ \unit{\per\cubic\cm}. In this material, the N shallow donor can compensate the shallow B acceptor and the acceptor levels of V\textsubscript{C} to pin the Fermi level at the N shallow donor level. 

Sample B is high-purity semi-insulating (HPSI) 4H-SiC material from Cree, used for stabilizing the charge state of single V\textsuperscript{4+} emitters. In this HPSI material, the residual N donors and B acceptors have comparable concentrations, in the low $10^{15}$ \unit{\per\cubic\cm} range. The Fermi level is located at the $\left(0|-\right)$ acceptor level of the divacancy at $\sim V_B + 2.05$ eV if the concentration of the N donor is larger than that of the B acceptor, or at the $\left(+|0\right)$ donor level of V\textsubscript{C} at $\sim E_V + 1.75$ eV if B has a higher concentration and can compensate the N donor (see  Fig. \,\ref{fig:3} (a)). In both cases, the neutral charge state V\textsuperscript{4+} is stable since the Fermi level lies between the $\left(+|0\right)$ and $\left(0|-\right)$ levels of V. In the former case, electrons trapped at the $\left(0|-\right)$ acceptor level of the divacancy (V\textsubscript{Si} V\textsubscript{C}) may still have weak influence on the PL decay of single V\textsuperscript{4+} emitters.

In both samples, vanadium ions were implanted at an energy of \num{100} \unit{\keV}, corresponding to a depth of about \num{60} \unit{\nm}. An implantation dose of \num{10}$^{8}$ \unit{\per\square\cm} creates a concentration sufficiently low to enable single-emitter studies. To repair the lattice damage created by the implantation process, the sample was annealed at 1400 $^{\circ}$C in Ar atmosphere for 30 minutes. The annealing was performed without a C-cap layer and no noticeable degradation of the surface morphology was observed.

\textbf{Optical measurements}. 
All measurements were performed with the sample at \num{4.3} \unit{\K}, mounted in a closed-cycle cryostat (Montana Cryostation s100) with external shroud customised to bring an external neodymium permanent magnet at a distance down to \num{20} \unit{\mm} from the sample. Optical measurements were performed a with a home-made confocal microscopy setup described in detail in Supplementary Note 1. We use two laser excitations, a narrowband (\num{20} \unit{\kHz}) tunable telecom laser (in the range \num{1270} \unit{\nm} - \num{1350} \unit{\nm}), in resonance with the ZPL of the V $\alpha$-line \cite{spindlberger_optical_2019}) and a green repump laser (\num{520} \unit{\nm}) to compensate for laser-induced ionisation \cite{wolfowicz_vanadium_2020}. A series of long pass filters in the detection path, allow us to filter out the excitation laser and collect only the phonon sideband of the emission (\num{1300} \unit{\nm} - \num{1600} \unit{\nm}), which is finally detected by a superconducting nanowire single-photon detector (Single Quantum EOS, detection efficiency \num{85}\% along one linear polarisation). See Supplementary Note 1 for a detailed description of the setup and the experimental measurements.

\textbf{Automated detection of quantum emitters}. 
In order to retrieve spectra for each of the spots, we took a sequence of $N$ maps at different detunings $\lbrace f_i \rbrace, i= 1..N$. PL spots associated are detected in each PLE maps by convolving the image with a Gaussian function of a similar width to the confocal spot size and detecting local maxima. The number of photon counts in each spot is then computed integrating the signal over a square as large as the confocal width, for each spot $k$. Spots with centres closer to each other than the spot radius are merged. For each spot $k$, we record the total number of photon counts in each of the maps at different detunings, retrieving the spectrum $P_k (f_i)$ and fit it with a linear combination of Gaussian functions. The centres of all the Gaussians are then used in the histograms in Fig. \ref{fig:inhomogeneous_broadening}(c).

\section*{Data availability}

The raw experimental data generated in this study have been deposited in a Zenodo database \cite{cilibrizzi_2023_10066455}.

\begin{acknowledgments}

We thank Caspar van der Wal, Carmem Gilardoni, Roland Nagy, Margherita Mazzera, Erik Gauger, Yoann Altmann and Brian Gerardot for helpful discussions and comments on our manuscript. 

This work is supported by the European Commission project QuanTELCO (grant agreement No 862721; M.T., C. Bonato, G.B., N.T.S.), the Engineering and Physical Sciences Research Council (EP/S000550/1; C. Bonato), the Leverhulme Trust (RPG-2019-388; C. Bonato), the Austrian Research Promotion Agency project QSense4Power (FFG 877615; M. T.), the Swedish Research Council (VR:2020-05444; J.U.H.) and the Knut and Alice Wallenberg Foundation (KAW 2018.0071; J.U.H, N.T. S.).

\end{acknowledgments}

\section*{Author contributions}
C. Bonato and M. T. conceived the experiments and supervised the project. P. C., M.J.A., C. Bekker, D. W. and C. Bonato constructed the experimental setup. P. C. and M.J.A. performed the experimental measurements. J.U.H designed and performed epitaxial growth, J.U.H and M.G. performed annealing. B. T. and G. B. simulated the expected vanadium spectra and provided general theoretical input relevant for data analysis. P.C., M. J. A., B. T., N. T. S., I. G. I., T. A., P. K. M. T. and C. Bonato analysed the data with input from all co-authors. P. C., M. J. A., B. T., N. T. S., M. T. and C. Bonato prepared the manuscript, with input from all co-authors.

\section*{Competing interests}
The authors declare no competing interests.

\clearpage

\renewcommand{\thepage}{S\arabic{page}}
\renewcommand{\thesection}{S\arabic{section}}
\renewcommand{\thetable}{S\arabic{table}}
\renewcommand{\thefigure}{S\arabic{figure}}
\renewcommand{\figurename}{Supplementary Figure}
\setcounter{page}{1}
\setcounter{figure}{0}

\section*{Supplementary Information}

\section*{Supplementary Note 1: Details of the Experimental Setup}
\label{SI:Setup}
The measurements described in the main text were performed in a home-made setup shown in Supplementary Figure \ref{fig:Setup}, based on a confocal microscope. 
\begin{figure*}[!ht]
\centering
\includegraphics[width=1\textwidth]{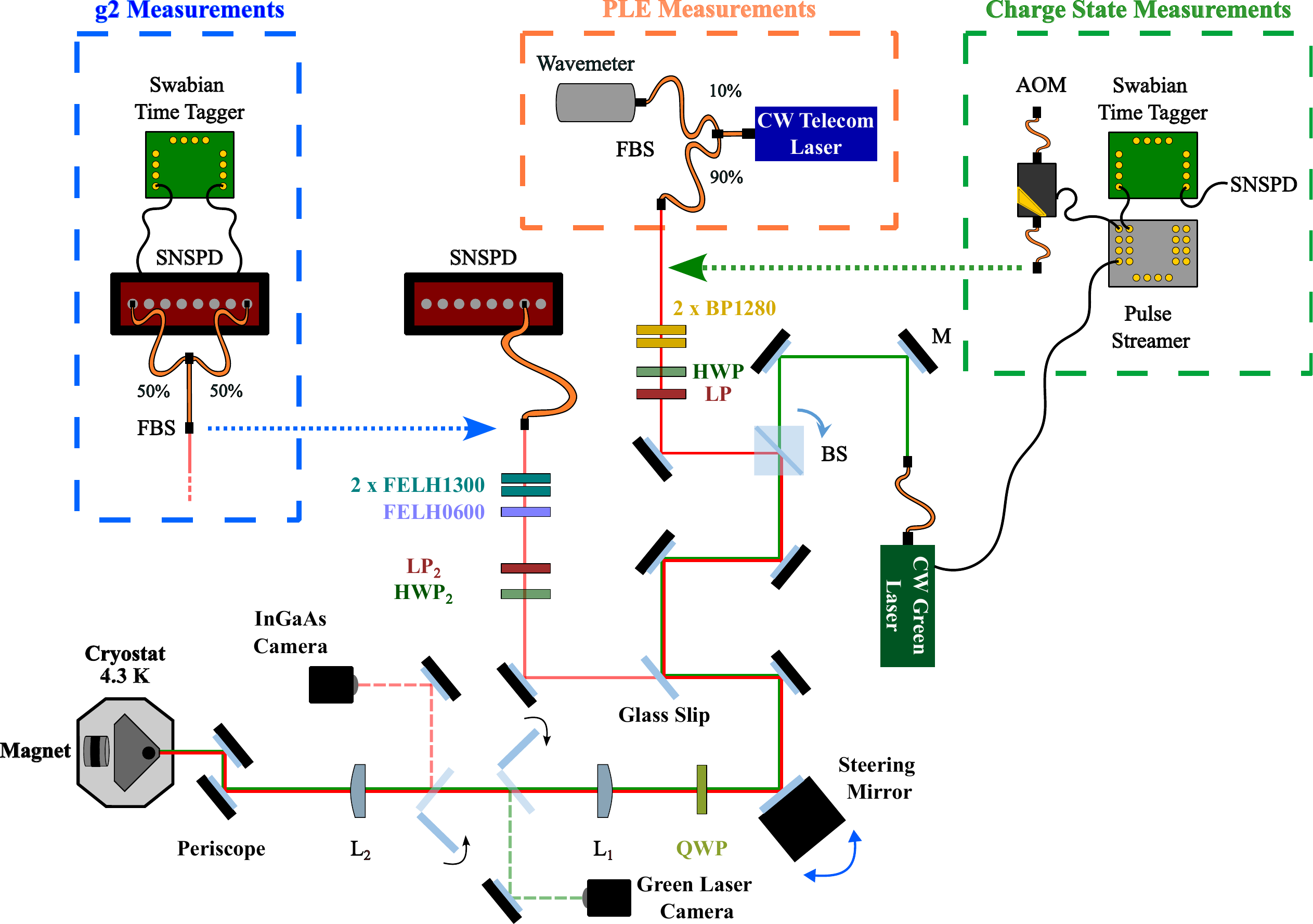}
\caption{\textbf{Sketch of the experimental setup.} The optical components, when not explicitly specified, are labelled as follows: \textbf{M} = Mirror, \textbf{BP1280} = Band Pass Filter, \textbf{HWP} = Half-Wave Plate, \textbf{LP} = Linear Polarizer, \textbf{BS} = Beam Splitter, \textbf{QWP} = Quarter-Wave Plate, \textbf{L1, L2} = Achromatic Doublet Lenses, \textbf{FELH0600} = Hard-Coated Long Pass Filter (Thorlabs), \textbf{FELH1300} = Hard-Coated Long Pass Filter (Thorlabs), \textbf{FBS} = Fiber Beam Splitter, \textbf{SNSPD} = Superconducting Nanowire Single-Photon Detector, \textbf{AOM} = Acousto-Optic Modulator. The small arrows close to some optical components indicate a flip mount.}
\label{fig:Setup}
\end{figure*}

The sample is mounted in a closed-cycle cryostat (Montana Cryostation S100). The cryostat comprises a 100x microscope objective (Olympus LCPLN100XIR, NA 0.85), tailored for near-infrared measurements, and kept at room temperature by a feedback-controlled heater. The sample is mounted on top of a stack of Attocube positioners (ANPx101). The cryostat shroud is customised to allow for an external permanent magnet to reach within 20 mm from the sample. A cylindrical (two-inch diameter, two-inch thick) Neodymium magnet, external to the cryostat (Supplementary Figure \ref{fig:Setup}), is used in our experiments to generate magnetic fields up to $\sim$ 100 mT, roughly aligned along the SiC c-axis. We use a commercially available gaussmeter (Tunkia TD8620) to characterize the intensity of the magnetic field at different distances from the two-inch cylindrical magnet.

The $\alpha$ zero-phonon line of vanadium in SiC is optically excited by two lasers: (1) a wavelength-tunable CW telecom diode laser (Toptica DL Pro, 1270 nm-1350 nm) tuned in resonance with the zero-phonon line (ZPL) of the GS1-ES1 transition ($\sim$1278.46 nm) and (2) a green diode laser (Thorlabs PL520, 520 nm), which is used as repump laser to stabilize the V$^{4+}$ charge state of the vanadium impurity. Both lasers are coupled into their respective excitation paths via two single-mode fibres (i.e., Thorlabs P3-980A-FC-2 and Thorlabs P3-460B-FC-2), and they are collimated by means of aspheric lenses. The excitation power and the polarization of the telecom laser are controlled by a combination of a half-wave plate and a linear polarizer (respectively, "HWP" and "LP" in Supplementary Figure \ref{fig:Setup}). The frequency of the telecom laser is monitored and stabilized by means of a Fizeau-Interferometer Wavemeter (HighFinesse, WS7).

A steering mirror (Newport FSM-300-NM) followed by a 4f optical system (i.e., a telescope), composed by two plano-convex lenses, L$_1$ (Newport - PAC35AR.16) and L$_2$ (Newport - PAC13AR.16) in Supplementary Figure \ref{fig:Setup}, allow the scanning of the excitation spot across the sample. In particular, the 4f telescope has the double function of expanding the collimated excitations beams (by a factor of 1.3), to fill the back aperture of the objective, while ensuring that both excitations lasers enter the back aperture of the objective through the centre, when the beams are stirred by the mirror.

The photoluminescence (PL), collected through the same excitation objective, is coupled into a single mode fibre (Thorlabs SMF28) and detected by a superconducting nanowire single-photon detector (SNSPD; Single Quantum EOS). We use a combination of iris diaphragms and two cameras, one for the telecom laser (Hamamatsu C10633, "InGaAs camera" in Supplementary Figure \ref{fig:Setup}) and one for the repump laser ("Green Laser Camera" in Supplementary Figure \ref{fig:Setup}), to align both the optical excitation and collection paths parallel to the c-axis of the sample. 
 
 We optically excite the V defects in resonance with the ZPL and collect only the phonon sidebands (PSB), using a glass microscope slide ("glass slip" in Supplementary Figure \ref{fig:Setup}) that reflects $\sim$3$\%$ of the excitation laser and a combination of spectral filters in both the excitation and detection paths. In particular, the telecom excitation laser is spectrally filtered using two 12 nm-FWHM bandpass filters, with 1280 nm central wavelength (Knightoptical, 1280DIB25, referred to as "BP1280" in Supplementary Figure \ref{fig:Setup}). An additional combination of longpass filters (i.e., Thorlabs FELH1300 and FELH0600 in Supplementary Figure \ref{fig:Setup}) in the detection path, ensures the suppression of both excitation lasers, and the collection of only photons with wavelength $\geq$ 1300 nm (i.e., corresponding to the PSB of the V defects \cite{wolfowicz_vanadium_2020}).\\ 

\textbf{{\boldmath $g^{(2)}(\tau)$} Measurements} \\
To perform $g^{(2)}(\tau)$ autocorrelation measurements, we use a Hanbury-Brown and Twiss setup. The PL emission from the defects is split by a fiber beam splitter ("FBS", Thorlabs TW1300R5F1), with 50:50 split ratio, as indicated in the "g2 measurements" blue box in Supplementary Figure \ref{fig:Setup}, and then detected by two SNSPDs. We use two channels of the TimeTagger (Swabian Instruments), to record and time-correlate the arrival times of the PL photons emitted by the V emitters. To test and characterize the photon arrival time delay of our experimental setup, we performed $g^{(2)}(\tau)$ measurements of a known single emitter source (NV centre sample). 



\textbf{Pulsed Measurements} \\
To study the charge state dynamics of V defects we perform time resolved PL measurements, by pulsing both the telecom and the repump lasers (depending on the experiments, as indicated in the manuscript) and time tagging the emitted PL photons. The telecom laser is pulsed by a fibre-coupled acousto-optic modulator (AOM, Aerodiode - 1310-AOM-2), while the green laser is directly pulsed through its Thorlabs LDC205C controller. All control pulse sequences, as well as trigger signals for the time-tagger, are generated by a waveform generator (Pulse Streamer 8/2, Swabian Instruments). Finally, to time resolve the PL signal emitted by the V defects, we use a time tagger (Swabian Instruments) directly connected to the SNSPD and to the Pulse Streamer.


\section*{Supplementary Note 2: Additional details on charge state dynamics}
\label{SI:charge}
\subsection{Charge transition levels}
The single acceptor level (0|-) of V in 4H-SiC has been determined to be at \num{0.97} \unit{\eV} below the conduction band (CB) ($E_C$ - \num{0.97} \unit{\eV}), by monitoring the transmutation of implanted radioactive $^{51}$Cr isotope to $^{51}$V using deep level transient spectroscopy \cite{achtziger_band_1997, achtziger_band-gap_1998}. 
The energy of the V single donor level (+|0), on the other hand, has not been unequivocally assigned in the literature. Initial studies 
 using optical admittance spectroscopy (OAS) in 6H-SiC \cite{evwaraye_optical_1996} and photo-excitation electron paramagnetic resonance (photo-EPR) in both 6H- and 4H-SiC \cite{baur_transition_1997}, assigned an energy of $\sim$ \num{1.6} \unit{\eV} above the valence band (VB) (EV + 1.6 eV) to the donor level (+|0). Later studies of V-doped semi-insulating 4H-SiC \cite{mitchel_fermi_1999}, found an activation energy of $\sim$ \num{1.6} \unit{\eV} from the temperature dependence of free carriers and assigned to the (+|0) donor level of V. This energy was assigned as the energy distance of the (+|0) level from the CB, $E_C$ - \num{1.6} \unit{\eV} \cite{mitchel_fermi_1999}. However, a deep donor level can compensate shallow acceptors in p-type materials but not shallow donors and, therefore, cannot pin the Fermi level in n-type materials. Thus, the activation energy measured in \cite{mitchel_fermi_1999} should be related to the VB and the (+|0) level of V should be at $\sim$ $E_V$ + \num{1.6} \unit{\eV}, in agreement with OAS \cite{evwaraye_optical_1996} and photo-EPR results \cite{baur_transition_1997}, and the Langer-Heinrich rule \cite{Langer_Impurities_Levels_1985}.

\subsection{Details on dynamics of activating and deactivating V\textsuperscript{4+}}
Most probably, the resonant laser (\num{1278.84} \unit{\nm} or \num{0.9695} \unit{\eV}) does not excite electrons from the (0|-) level of V (at E\textsubscript{C} – \num{0.97} \unit{\eV}) to the CB. Indeed, the value of E\textsubscript{C} – \num{0.97} \unit{\eV} for the (0|-) acceptor level of V quoted in literature \cite{achtziger_band_1997, achtziger_band-gap_1998} is based on Deep Level Transient Spectroscopy (DLTS) and obtained at rather high temperature above 400 degrees, where the bandgap is at least few tens of \unit{\meV} ($\sim$ \num{60} \unit{\meV} \cite{choyke_1969}) smaller than at cryogenic temperatures. Assuming that the (0|-) level is pinned to the VB one estimates that the V acceptor level is about \num{1.03} \unit{\eV} below the CB at cryogenic temperature (E\textsubscript{C} – \num{1.03} \unit{\eV}), hence \num{0.97} \unit{\eV} optical excitation cannot ionise the V acceptor level. However, as mentioned in the main text, the \num{0.97} \unit{\eV} excitation efficiently removes electrons from the higher-lying (0|-) and (-|2-) levels of V\textsubscript{C} (at $\sim$ \num{0.5} \unit{eV} and $\sim$ \num{0.7} \unit{eV} below the CB, respectively) and ionises neutral N donors. With higher power of the infrared laser more electrons are generated in the CB, which enhances the rate of capturing electrons to the acceptor level of V, leading to a higher PL quenching rate. The green repump laser also can ionize shallow donor and acceptor (VC) levels, but it also ionizes V\textsuperscript{3+} to create the bright state V\textsuperscript{4+}.

Thus, the decaying process of the V\textsuperscript{4+} PL reflects the depletion of the free-electron population in the CB created by either the green or the resonant infrared excitation. This notion is supported by PL experiments with pulsed green and infrared excitations with a delay $\tau$ between the pulses shown in Figure 4(c) in the main text. The maximum PL intensity decays exponentially with increasing delay $\tau$ because the population of the neutral V charge state decays exponentially together with the free-electron concentration.

The decay rate is sample dependent, as shown by comparison of the decay times of samples A and B (Figure 4(c) in the main text), because the free electron concentration generated by either the excitation or the re-pump is very different in these two samples, owing to different Fermi level positions, as discussed in the main text.

\subsection{Experimental PL decay curves}

In this section, we report some of the experimental PL decay curves, acquired to investigate the ionisation of the V centre in Fig 4 in the main text.

As described in the main text, we prepared the V\textsuperscript{4+} state with a \num{300} \unit{\ms} green laser pulse and detected the PL decay as a function of different (telecom) resonant excitation powers. Green pulses are repeated every \num{700} \unit{\ms}, a time sufficient for the detected photon intensity to be completely decayed. 

In Supplementary Figure \ref{fig:3_decay_curves} (a), we report two of the decay curves acquired with the same excitation power (\num{6} \unit{\uW}) on sample A (violet) and sample B (orange), respectively. 
In Supplementary Figure \ref{fig:3_decay_curves}(b), we report PL decay curves at different excitation powers (\num{2} \unit{\uW}, \num{8} \unit{\uW} and \num{25} \unit{\uW}, respectively) for sample B.
To obtain the data shown in Figure 4(b) of the main text, we fitted the decay curves acquired at different excitation powers with a single exponential function, as shown in Supplementary Figure \ref{fig:3_decay_curves}(b). The decay constants extracted from the fits were plotted as a function of the (telecom) resonant excitation power in Figure 4(b) of the main text.
As described in the manuscript, a similar procedure allowed us to extract the decay constant shown in Figure 4(c) of the main text, as a function of the delay time ($\tau$). Specifically, we measured the PL decays of the same V centre at different green and telecom laser pulse delays. The different delays correspond to time intervals when no illumination is present on the sample. We then fit the decays at different delays ($\tau$) using a single exponential, and plot the peak intensity, extracted from the previous fits, as a function of the delay time ($\tau$) (see Figure 4 (c) of the main text). For sample B in Fig. 4(c) of the main text, we repeated the experiment twice, and plotted the average of the two measurements. Error bars in Figs. 4(a) and 4(b) of the main text, correspond to the standard deviations extracted from the fits, calculated by taking the square root of the diagonal elements of the covariance matrix obtained during curve fitting.
\begin{figure*}[!hbtp]
\includegraphics[width=0.9\textwidth]{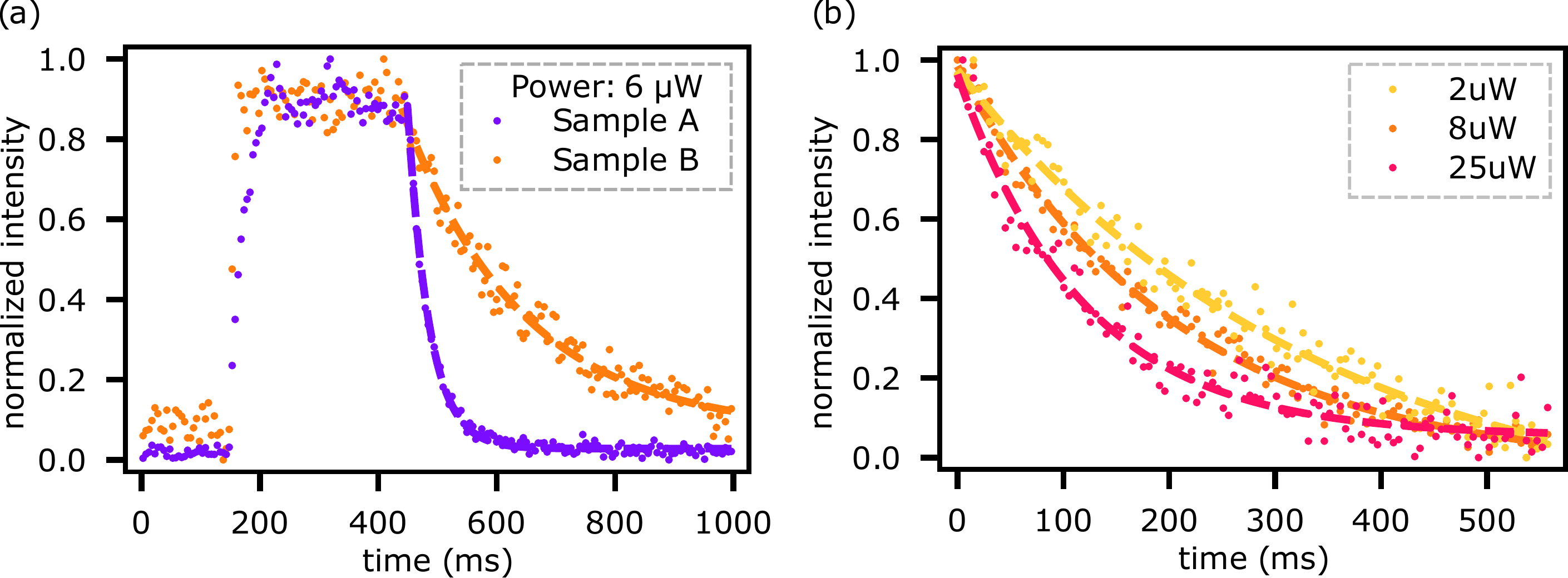}
\caption{\textbf{Charge state measurement details.} (a) Full decay curves and fittings of the V$^{4+}$ charge state acquired under a \num{300} \unit{\ms} green laser pulse and CW telecom excitation (see inset in Figure 4(b) of the main text for details on the excitation scheme). The telecom excitation power is \num{6} \unit{\micro\watt} for both sample A (violet) and sample B (orange). (b) Decay curves and fittings of the same PL spot in sample B at different excitation powers (see legend). Each of the curves is fit by a single exponential with a different decay constant.}
\label{fig:3_decay_curves}
\end{figure*}

\section*{Supplementary Note 3: Summary of the V centre electronic structure}
\label{SI:electronic_structure}
\begin{table}
  \caption{\label{tab:par}
    Literature parameters for the relevant KDs for the Vanadium \(\alpha\) defect in 4H-SiC.
    For the ground states parameters from \cite{tissot-hyperfine-2021}, which are based on \cite{wolfowicz_vanadium_2020} are used and for the ES parameters from \cite{astner_vanadium_2022} are used.
    The signs of $a^{zz}_k$ are chosen to best describe the data.
    All remaining components of the $g$- and hyperfine-tensors are $0$.
  }
  \begin{ruledtabular}
    \begin{tabular}{l r r r r}
          KD \(k\) & \(g_{k}^z\) & \(a_{k}^{zz}/h\) (MHz) & \(a_{k}^{xx,yy}/h\) (MHz) & \(a_{k}^{xz}/h\) (MHz) \\
\hline
   $g$ (\(1,\Gamma_{4}\))  & 1.748 & 232 & 165 & 0 \\
   $e$ (\(2,\Gamma_{5/6}\)) & 2.24 & 213 & 0 & 75 \\
    \end{tabular}
  \end{ruledtabular}
\end{table}

Combining the theory developed in \cite{tissot-electronic-2021,tissot-hyperfine-2021,tissot_nuclear_2022}, we describe the structure of the optical transitions for V\textsuperscript{4+} using the lowest ground and excited state Kramers doublets (KDs).
To this end, we model the energies of the KDs using the KD Hamiltonians
\begin{align}
  \label{eq:HKD}
  H_k = & E^k
          + \frac{1}{2} \mu_B \vec{B} \mathbf{g}_{k} \vec{\sigma}_{k}
          + \frac{1}{2} \vec{\sigma}_{k} \mathbf{A}_k \vec{I}
          + \mu_N g_N \vec{B}\cdot \vec{I} ,
\end{align}
where $k$ labels the KDs, with the zero-field energies $E_k$ of the $k$-th KD  (in MHz),
the Bohr magneton $\mu_B$ and nuclear magneton $\mu_N$  (in MHz$/$Gauss),
the nuclear $g$-factor $g_N$,
 the strongly KD dependent $g$-tensor $\mathbf{g}_k$,
as well as the hyperfine tensor $\mathbf{A}_k$, see Supplementary Table ~\ref{tab:par}.
Coupling to an electric field with a frequency in the vicinity of the crystal field splitting
\(\Delta_{\mathrm{cr}} = E^{\mathrm{e}} - E^{\mathrm{g}}\)
can be described using the Hamiltonian
\begin{align}
  \label{eq:Hd_pol}
  H_d^{\sigma} \approx \mathcal{E}(t) \epsilon \ket{e,-\sigma}\bra{g,-\sigma}
  + \mathrm{h.c.}
\end{align}
where \(\sigma = \pm = \uparrow,\downarrow\)  describes the pseudo-spin of the KDs as-well-as the polarization of the electric field with amplitude \(\mathcal{E}(t)\) within the rotating wave approximation.
Here we only take the dominating part of the transition that conserves the spin into account
and thereby neglect the much weaker spin-flipping transition possible due to the spin-orbit coupling.
[The driving field should fulfil \(|\epsilon \mathcal{E}| \ll \min_i|\Delta_{\mathrm{so}}^i|\) for the RWA]

To then model the spectra we assume that each of the hyperfine transitions has the same width \(\gamma\).
We therefore quantify the transition intensity of each of the hyperfine transitions as
\begin{align}
  \label{eq:model}
  a \frac{\gamma |\langle g,\sigma_g,m_g \Vert {g,\sigma} \rangle \langle {e,\sigma} \Vert e, \sigma_e, m_e \rangle|^2}{4(\omega-E_{\sigma_e, m_e}^e+E_{\sigma_g,m_g}^g)^2 + \gamma^2}
\end{align}
where we use \(a\) as a fit parameter independent of the hyperfine states (but depending on \(B\)).
The hyperfine states are labeled according to their main spin \(\sigma_k\) and nuclear contribution \(m_k\)
but are made up of different spin, orbital and nuclear states.
Analytic expressions for the states and energies can be found in \cite{tissot-hyperfine-2021}.

To fit the data of Figure 2 of the main text, we sum all the allowed transitions for a given polarization \(\sigma\) and add a global offset \(o\) (dependent on \(B\)),
reusing existing hyperfine parameters \cite{tissot_nuclear_2022,astner_vanadium_2022}
the remaining parameters to determine the hyperfine linewidth \(\gamma\), the crystal splitting \(\Delta_{cr}\) (which can vary from defect to defect),
as well as the magnetic field dependent fit parameters \(a(B),o(B)\).
Using a least squares fit weighted by the inverse Poisson error of the data, we find
$\gamma = 1038\pm7\,$ \unit{\MHz} and central transition frequency
 $\Delta_{cr} = 234425594 \pm 4\,$ \unit{\MHz}.

\begin{table}
  \caption{\label{tab:ao}
    Measurement setup dependent fit parameters. The unit of $a(B)$ is due to the unit of the Lorentzian being MHz.
  }
  \begin{ruledtabular}
    \begin{tabular}{l r r r}
          Parameter & 0\,$\mathrm{Gauss}$ & 600\,$\mathrm{Gauss}$ & 1000\,$\mathrm{Gauss}$ \\
\hline
Amplitude $a(B)$ \, (GHz/s) & 22.6$\pm$0.2 & 19.5$\pm$0.2 & 23.1$\pm$0.2 \\
Offset $o(B)$ \, (1/s) & 24.9$\pm$0.3 & 23.4$\pm$0.3 & 27.6$\pm$0.3
    \end{tabular}
  \end{ruledtabular}
\end{table}

We highlight that the narrowing of the hyperfine transitions suggests that the signs of the $a_k^{zz}$ with $k=e,g$ are the same
and that the measurement confirms the theory predicted polarization dependent selection rules.

\section*{Supplementary Note 4: Additional information on the inhomogeneous spectral distribution}
\label{SI:Inhom_Distrib_Table}

\setcounter{subsection}{0} 
\subsection {Comparison between different emitters}
In Supplementary Table \ref{table:review}, we compare the inhomogeneous distribution and lifetime-limited linewidth for different quantum emitters in different materials platforms. Given the goal of bringing any two quantum emitters into resonance, the important figure of merit here is the ratio $\eta$ between the inhomogeneous distribution of emitters and the linewidth. We take the ideal case of a lifetime-limited linewidth, computed from the optical lifetime $\tau$ as $1/(2 \pi \tau)$: a linewidth as close as possible to the lifetime limit is crucial to achieve high-visibility quantum interference. 


We include quantum emitters from different material platforms, including deep-level point defects in diamond, SiC, silicon and rare-earth ions in crystals. With an inhomogeneous distribution of $40-50$ \unit{\GHz}, the NV centre in diamond scores $\eta > 3000$, i.e. a lifetime-limited line of \num{13} \unit{\MHz} may have to be tuned across more than \num{3000} linewidths to bring any NV into resonance with any other NV. The silicon vacancy in diamond (SiV) and silicon carbide (V\textsubscript{Si}) achieves a better value for $\eta$, $\sim 100$ and $>500$, respectively. It's important to stress that SiV in diamond features first-order insensitivity to strain and electric fields due to inversion symmetry. We did not include the divacancy in SiC in the Supplementary Table, as we could not find an experimental value for its inhomogeneous broadening in the literature: given that its electronic structure and spin stare are identical to the NV centre in diamond, we however expect similar numbers. As we have seen, the inhomogeneous distribution for vanadium in SiC can be reduced from several \unit{\GHz} ($\eta>10^4$) in standard material to $\sim 100$ \unit{\MHz} ($\eta \sim 100$) in isotopically-enriched SiC. Recent work on the T centre in Si has demonstrated extremely narrow inhomogeneous broadening, down to \num{60} \unit{\MHz} \cite{simmons_T_centre_PRXQ}, in ensemble experiments ($\eta \sim 350$): the broadening is however much larger when creating single emitters by implantation ($\eta >5000$) \cite{simmons_single_T_NJP2021}. Rare-earth ions show outstanding coherence of their optical transitions, with inhomogeneous broadening down to the MHz regime, in some cases (e.g. ensemble \textsuperscript{153}Eu\textsuperscript{3+}:EuCl\textsubscript{3} $\cdot$ 6H\textsubscript{2}O in the Supplementary Table \ref{table:review}). Their long optical lifetimes, in the ms regime, with lifetime-limited linewidths in the few tens of Hz regime however results in quite high values for $\eta$.

 \begin{table*}[ht]
 \caption{Summary of inhomogeneous distribution for several spin-active quantum emitters}
 \begin{tblr}{
   @{}rrrrrX[r,valign=b
   ]X[r,valign=b]X[r,valign=b]@{}
 }
 \hline
 \hline
 centre & wavelength &
 lifetime  &  
 Fourier linewidth & 
 inhomogen. distr. & 
 $\eta$ \\
 \hline
 NV:diamond (single)& 637 nm &12 ns & 13 MHz & $\sim$40-50 GHz \cite{NV_centers_Strain_2019} & >3000 \\
 SiV:diamond (single)& 750 nm & 1.7 ns & 94 MHz & $\sim$10 GHz \cite{Rogers_Inhomogenous_Si_in_Diamond_2014, SiV_Diamond_Inhomogeneous_2016} & 107\\
 V\textsubscript{Si}:SiC (single) & 861 nm & 6 ns & 27 MHz & $\sim$15 GHz\cite{nagy_APL_2021} & 565\\
 V:SiC (single) & 1278 nm & 167 ns & 0.95 MHz & $>$10 GHz & >10,000 \\
 \textbf{V:\textsuperscript{28}Si\textsuperscript{12}C} (single) & \textbf{1278 nm} & \textbf{167 ns} & \textbf{0.95 MHz} & \textbf{$\sim$100 MHz} & \textbf{105} \\
Er\textsuperscript{3+}:Si (single) & 1536 nm & $\sim 0.186$ ms \cite{gritsch2022narrow} & $\sim$ 856 Hz & $\sim$0.5 GHz \cite{gritsch2022narrow}& $\sim 0.6 \cdot 10^6$ \\
 Er\textsuperscript{3+}:YSO (single) & 1550 nm & $\sim 11$ ms \cite{thomson_PRL_2018} & $\sim$ 14.5 Hz & $\sim$20 GHz \cite{thompson_SSRO_single_ions_science2020}& $>1 \cdot 10^9$ \\
 T centre:Si (single) & 1326 nm & 940 ns \cite{simmons_T_centre_PRXQ} & 169 kHz & $\sim$1 GHz \cite{simmons_single_T_NJP2021} & 5900 \\
 G centre:Si (single) & 1280 nm & $\sim$ 36 ns \cite{Redjem_G_center_2020} & 4.4 MHz & >100 GHz \cite{dreau_G_centre} & >27000 \\
 T centre:\textsuperscript{28}Si (ensemble) & 1326 nm & 940 ns \cite{simmons_T_centre_PRXQ} & 169 kHz & 60 MHz \cite{simmons_T_centre_PRXQ} & 355 \\
 G centre:\textsuperscript{28}Si (ensemble) & 1280 nm & $\sim$ 36ns \cite{Redjem_G_center_2020} & 4.4 MHz & 72 MHz  \cite{thewalt_PRB_linewidths_28Si} & 16 \\
 
 \textsuperscript{171}Yb\textsuperscript{3+}:Y\textsubscript{2}SiO\textsubscript{5} (ensemble) & $\sim$ 980 nm & $\sim$ 1ms \cite{afzelius_inhmg_broad_Yb_PRB2018} & $\sim$ 160 Hz & $\sim$ GHz \cite{afzelius_inhmg_broad_Yb_PRB2018} & $>6 \cdot 10^6$ \\
 \textsuperscript{153}Eu\textsuperscript{3+}:EuCl\textsubscript{3} $\cdot$ 6H\textsubscript{2}O (ensemble) & 579.7 nm & $\sim$ 2ms \cite{sellars_lifetime_Eu} & $\sim$ 80 Hz & 25 MHz \cite{sellars_inhmg_broad_Eu_PRL2016} & $>3 \cdot 10^5$ \\
 Eu\textsuperscript{3+} molecular crystal & 580.38 nm & $\sim$ 540 \unit{\micro \s} \cite{serrano_ultra-narrow_2022} & $\sim$ 295 Hz & $\sim$ 200 \unit{\MHz} \cite{serrano_ultra-narrow_2022} & $>6 \cdot 10^5$\\
 \hline
 \end{tblr}
 \label{table:review} 
 \end{table*}

\subsection{Experimental determination of inhomogeneous distributions}

\begin{figure*}[!ht]
\includegraphics[width=1\textwidth]{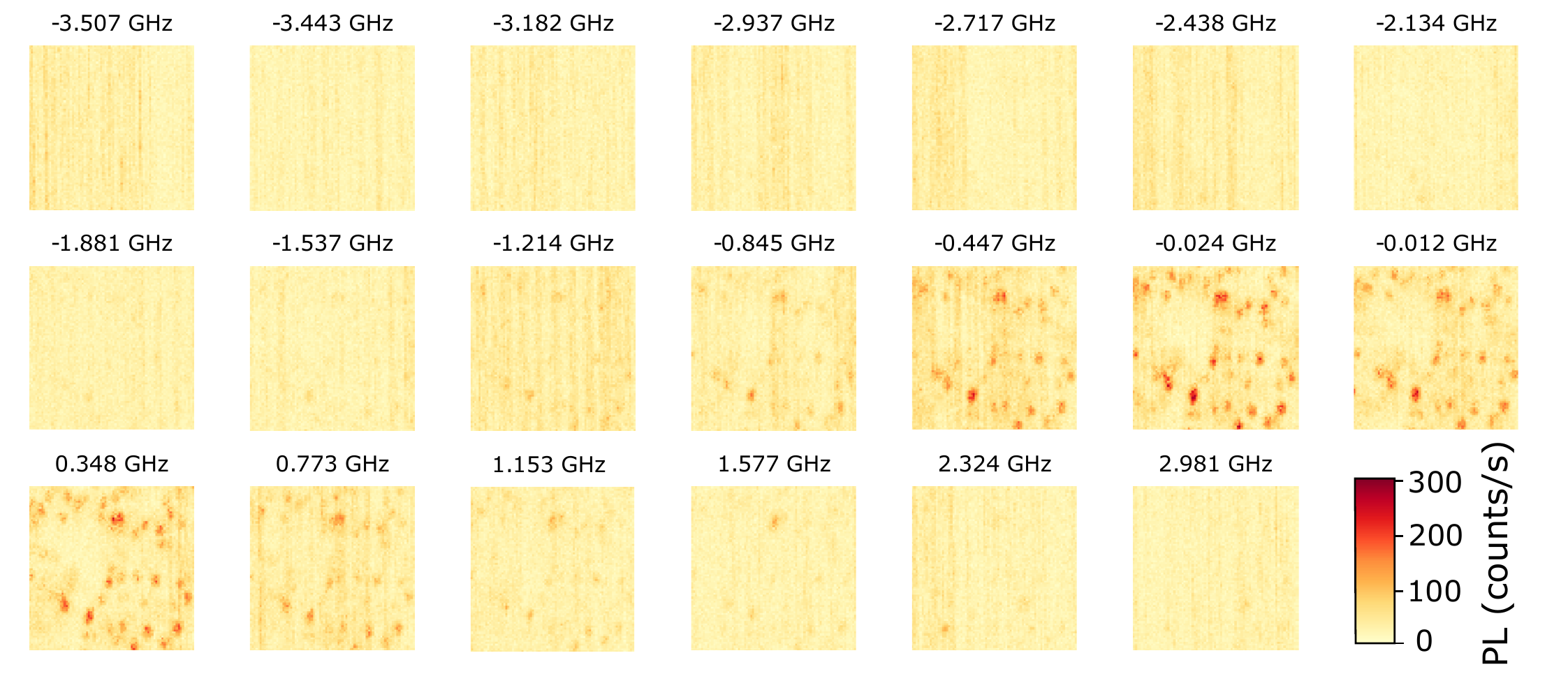}
\caption{\textbf{PLE maps (region A1, sample A).} Photoluminescence maps for different detunings of the telecom excitation laser for the isotopically-enriched sample. Each map shows a \SI{12}{\micro\meter} by \SI{12}{\micro\meter} area. For these measurements, we use an integration time of \SI{1}{\second} per step, an excitation power of \SI{14}{\micro\watt} for the repump laser, and \SI{4}{\micro\watt} for the telecom laser.}
\label{fig:maps_2june}
\end{figure*}

\begin{figure*}[!ht]
\includegraphics[width=0.9\textwidth]{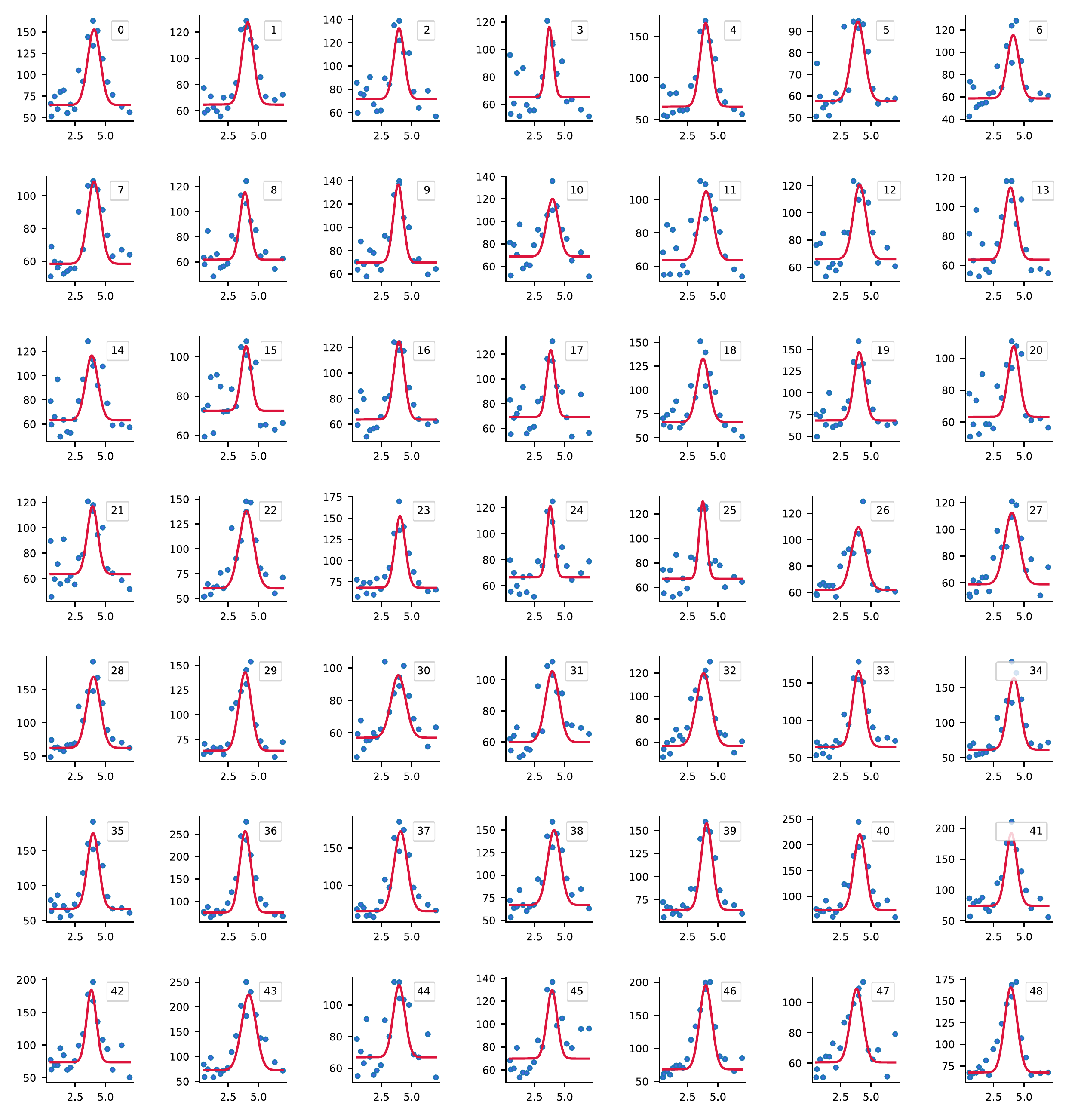}
\caption{\textbf{PLE spectra for each spot (region A1, sample A)}. Fitted spectra from \num{48} spots in Supplementary Figure \ref{fig:maps_2june}. For each sub-plot, the x-axis shows the excitation frequency in \unit{\GHz}, and the y-axis the number of detected counts per second.}
\label{fig:spectra_2june}
\end{figure*}

In this section, we present additional details about the estimation of the inhomogeneous distribution for emitters in the sample, described in Figure 3 of the main text. We performed a sequence of maps of the same area at different detunings of the excitation laser (see for example Supplementary Figure \ref{fig:maps_2june}). A custom python code automatically detects photoluminescence spots in each map and plots the integrated counts for each spot as a function of the detuning of the excitation laser (see for example Supplementary Figure \ref{fig:spectra_2june}). The photoluminescence peak for each spot is then fit with a Gaussian function. The single emitter nature of each spot was not checked, so some of the spots may correspond to multiple emitters.

We plot the sequence of maps and the corresponding spectral peaks for each spot (with the fit) for four regions (labelled A1-A4) in sample A (isotopically-enriched) and one region in sample B (natural abundance of isotopes), as follows:

\begin{itemize}
    \item region A1: PL maps in Supplementary Figure \ref{fig:maps_2june} and spectra in Supplementary Figure \ref{fig:spectra_2june}.
    
    \item region A2: PL maps in Supplementary Figure \ref{fig:maps_10june} and spectra in Supplementary Figure \ref{fig:spectra_10june}.
    
    \item region A3: PL maps in Supplementary Figure \ref{fig:maps_16sept} and spectra in Supplementary Figure \ref{fig:spectra_16sept}.
    
    \item region A4: PL maps in Supplementary Figure \ref{fig:maps_18dec} and spectra in Supplementary Figure \ref{fig:spectra_18dec}.

    \item sample B: PL maps in Supplementary Figure \ref{fig:map_NA1}, \ref{fig:map_NA2}, \ref{fig:map_NA3} and spectra in Supplementary Figure \ref{fig:spectra_NA1}, \ref{fig:spectra_NA2}.
\end{itemize}

For the four regions in sample A, the centres of the distributions of central frequencies vary slightly between the different regions: $f_{A1}$ = \num{227} $\pm$ \num{105} \unit{\MHz}, $f_{A2}$ = \num{-41} $\pm$ \num{109} \unit{\MHz}, $f_{A3}$ = \num{-25} $\pm$ \num{76} \unit{\MHz} and $f_{A4}$ = \num{712} $\pm$ \num{97} \unit{\MHz}. As evidenced by the standard deviations within each region, the central frequencies are very narrowly distributed over about \num{100} \unit{\MHz}. Note that the PLE maps for region A4 were acquired after a thermal cycle of warm-up to room temperature and cool-down to \num{4} \unit{\K}, which may have affected the sample strain. 
For the defects observed in the area D of the sample, we confirm the ultra-narrow inhomogenous broadening by high-resolution PLE spectroscopy of six different centres (Supplementary Figure \ref{fig:inhomogeneous_broadening_SI} b). The PLE spectroscopy measurements further corroborate the results of the automated statistical analysis.

\begin{figure*}[!ht]
\includegraphics[width=1\textwidth]{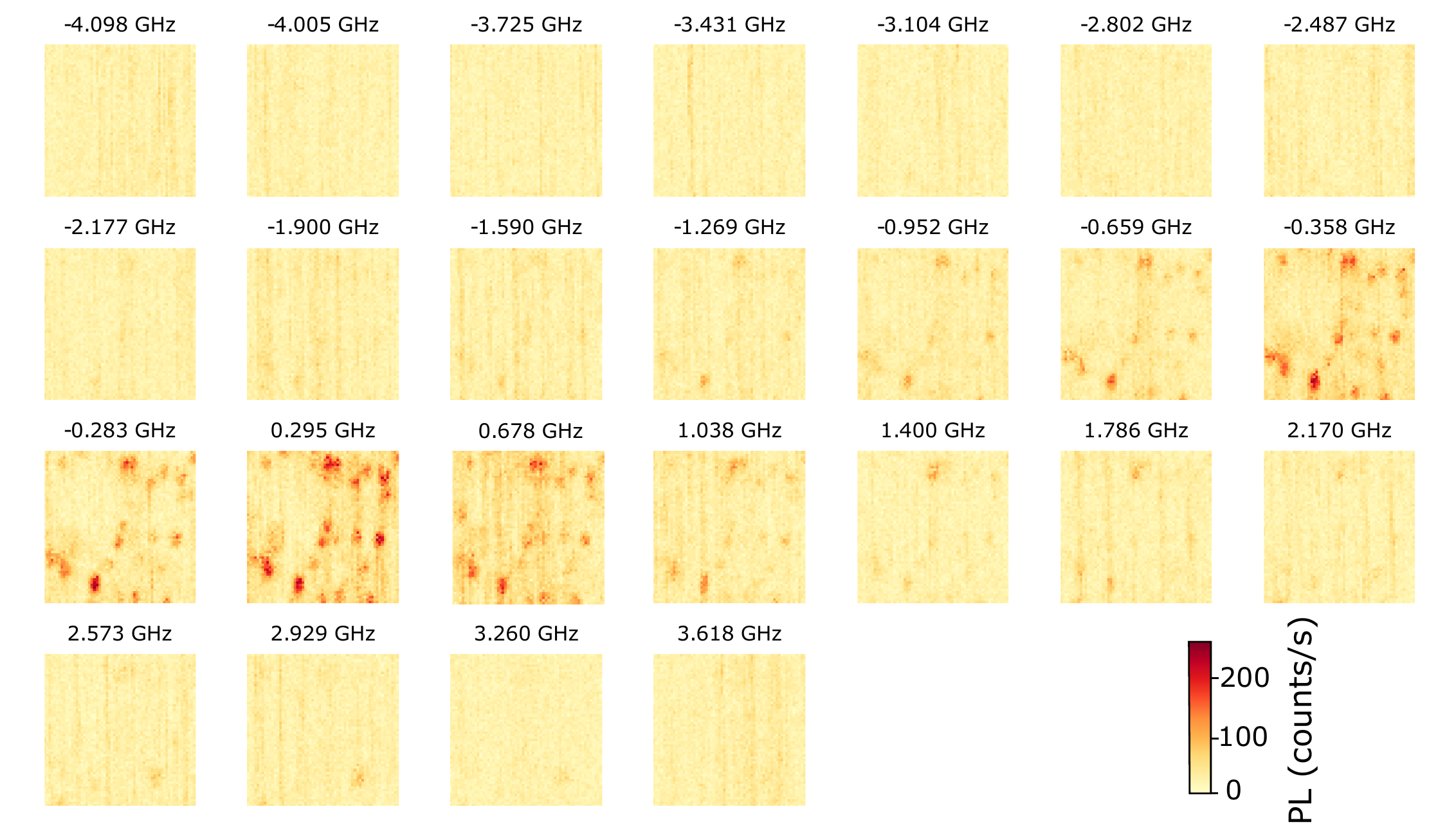}
\caption{\textbf{PLE maps (region A2, sample A).} Photoluminescence maps for different detunings of the excitation laser for the isotopically-enriched sample. Each map shows a \num{10} \unit{\micro\meter} by \num{10} \unit{\micro\meter} area. For these measurements, we use an integration time of \SI{1}{\second} per step, an excitation power of \SI{14}{\micro\watt} for the repump laser, and \SI{4}{\micro\watt} for the telecom laser.}
\label{fig:maps_10june}
\end{figure*}


\begin{figure*}[!ht]
\includegraphics[width=1\textwidth]{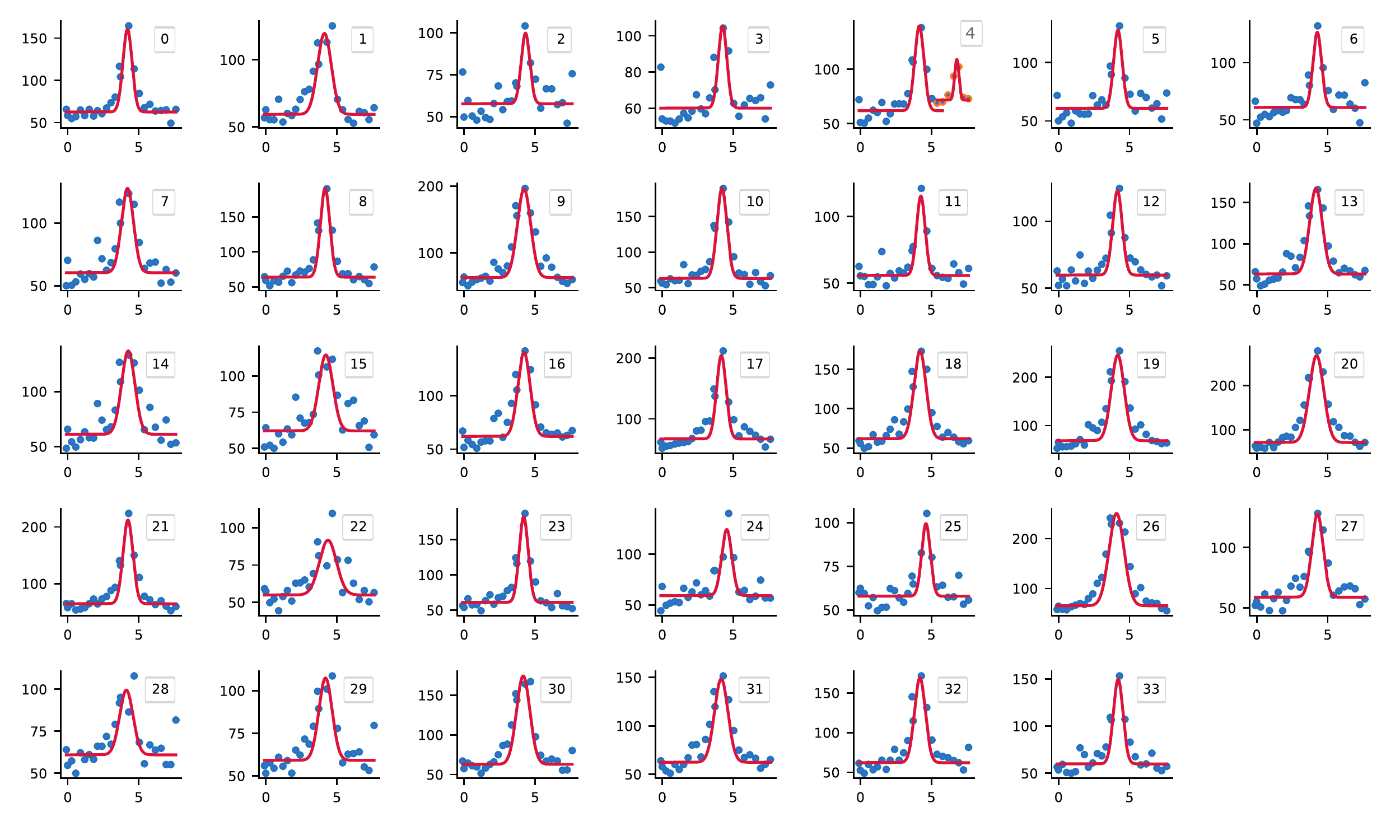}
\caption{\textbf{PLE spectra for each spot (region A2, sample A)}. Fitted spectra from \num{33} spots in Supplementary Figure \ref{fig:maps_10june}. For each sub-plot, the x-axis shows the excitation frequency in \unit{\GHz}, and the y-axis the number of detected counts per second.}
\label{fig:spectra_10june}
\end{figure*}

\begin{figure*}[!ht]
\includegraphics[width=1\textwidth]{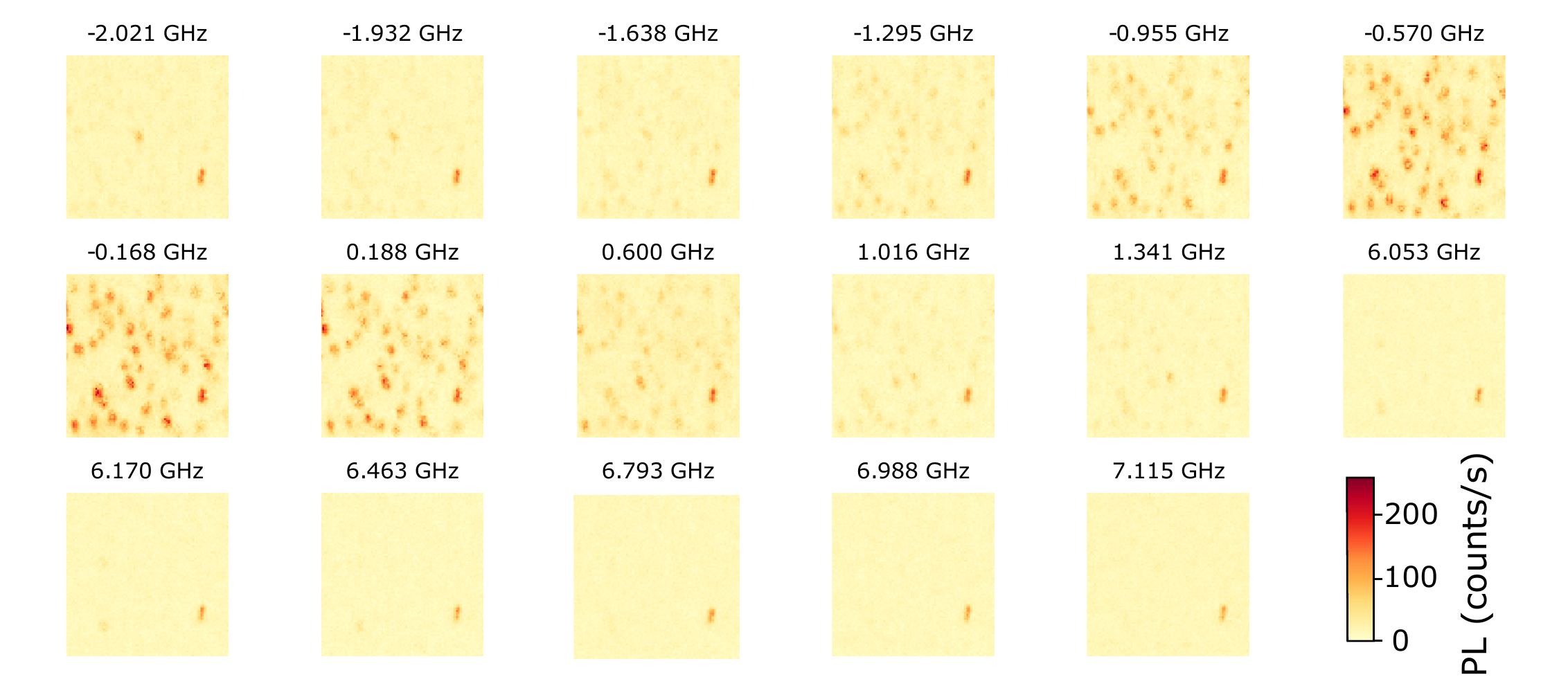}
\caption{\textbf{PLE maps (region A3, sample A).} Photoluminescence maps for different detunings of the telecom excitation laser for the isotopically-enriched sample. Each map shows a \SI{12}{\micro\meter} by \SI{12}{\micro\meter} area. For these measurements, we use an integration time of \SI{1}{\second} per step, an excitation power of \SI{14}{\micro\watt} for the repump laser, and \SI{2.2}{\micro\watt} for the telecom laser.} 
\label{fig:maps_16sept}
\end{figure*}

\begin{figure*}[!ht]
\includegraphics[width=0.9\textwidth]{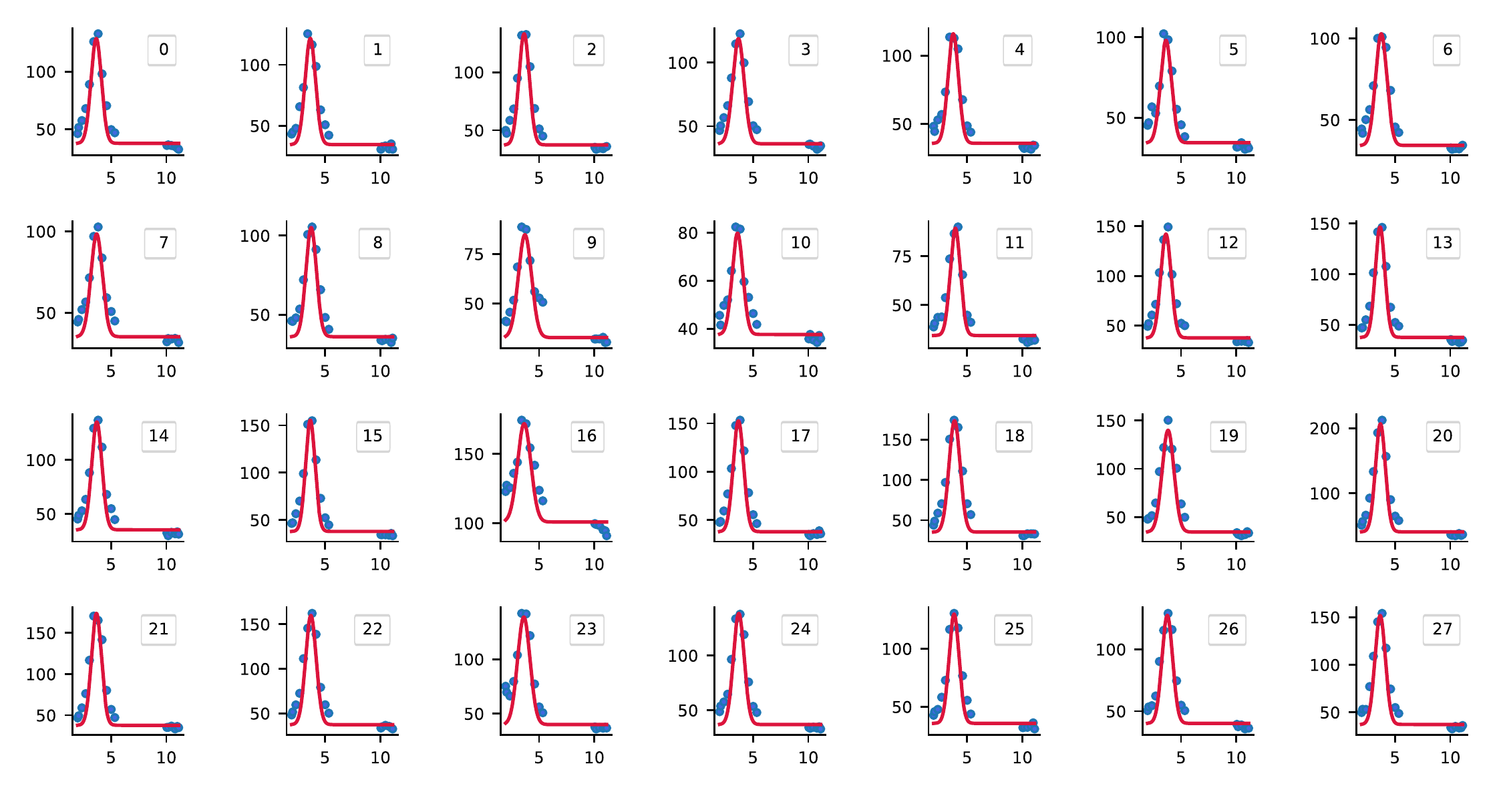}
\caption{\textbf{PLE spectra for each spot (region A3, sample A)}. Fitted spectra from \num{27} PL spots in Supplementary Figure \ref{fig:maps_16sept}. For each sub-plot, the x-axis shows the excitation frequency in \unit{\GHz}, and the y-axis the number of detected counts per second.}
\label{fig:spectra_16sept}
\end{figure*}

\begin{figure*}[!ht]
\includegraphics[width=1\textwidth]{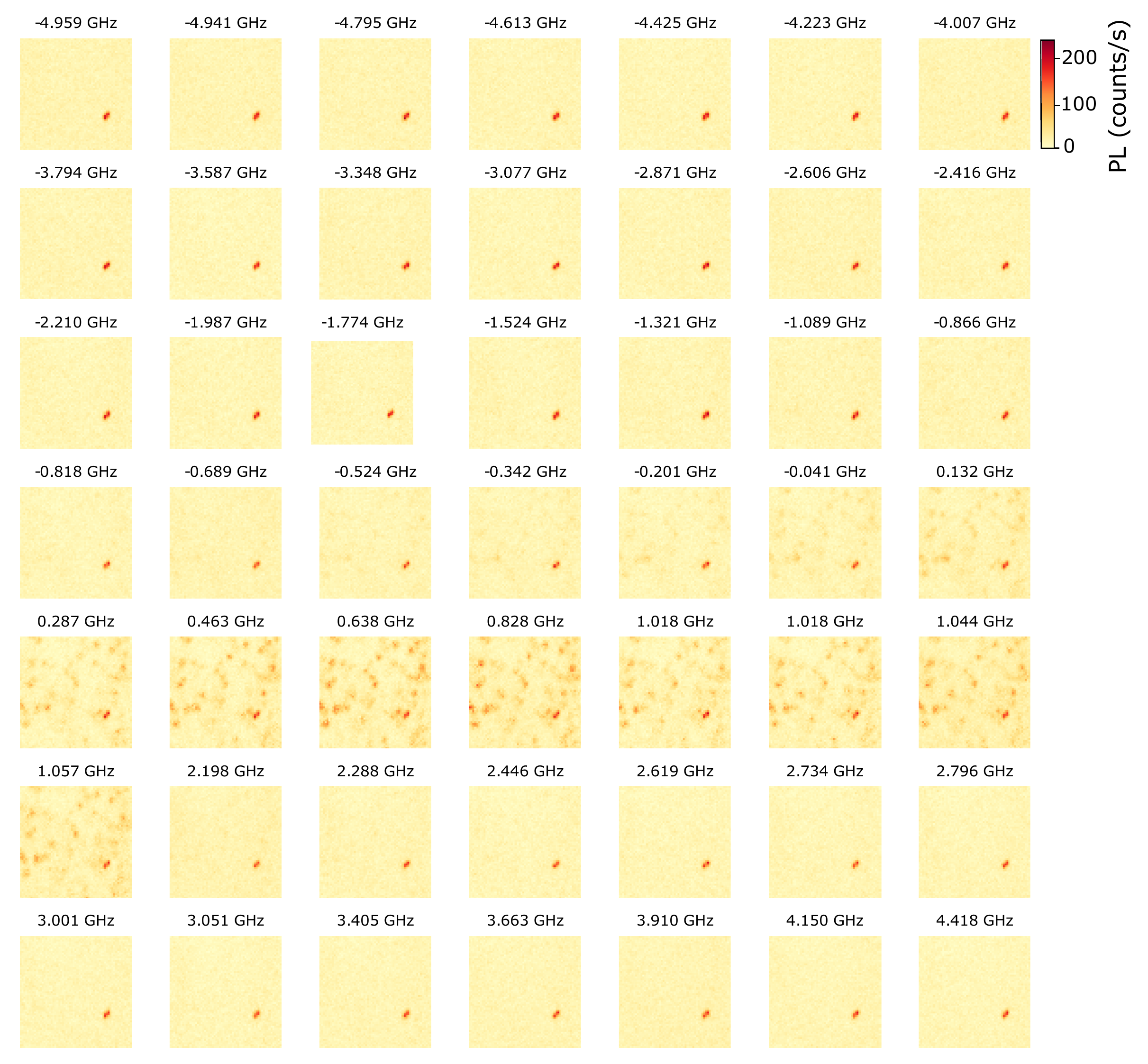}
\caption{\textbf{PLE maps (region A4, sample A).} Photoluminescence maps for different detunings of the telecom excitation laser for the isotopically-enriched sample. Each map shows a \SI{12}{\micro\meter} by \SI{12}{\micro\meter} area. For these measurements, we use an integration time of \SI{1}{\second} per step, an excitation power of \SI{14}{\micro\watt} for the repump laser, and \SI{3}{\micro\watt} for the telecom laser.}
\label{fig:maps_18dec}
\end{figure*}

\begin{figure*}[!ht]
\includegraphics[width=0.9\textwidth]{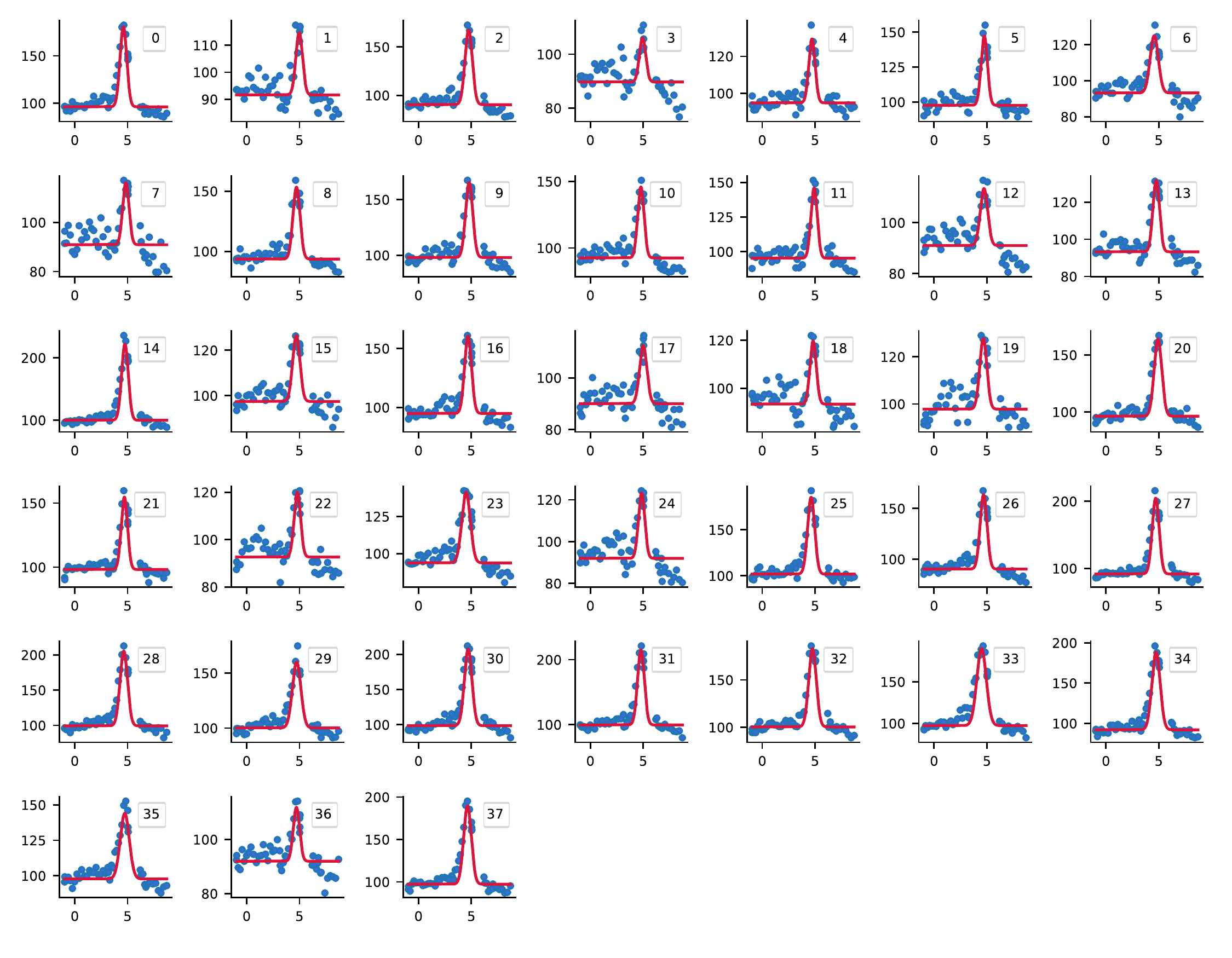}
\caption{\textbf{PLE spectra for each spot (region A4, sample A)}. Fitted spectra from \num{37} PL spots in Supplementary Figure \ref{fig:maps_18dec}. For each sub-plot, the x-axis shows the excitation frequency in \unit{\GHz}, and the y-axis the number of detected counts per second.}
\label{fig:spectra_18dec}
\end{figure*}

\begin{figure*}[!ht]
\includegraphics[width=0.9\textwidth]{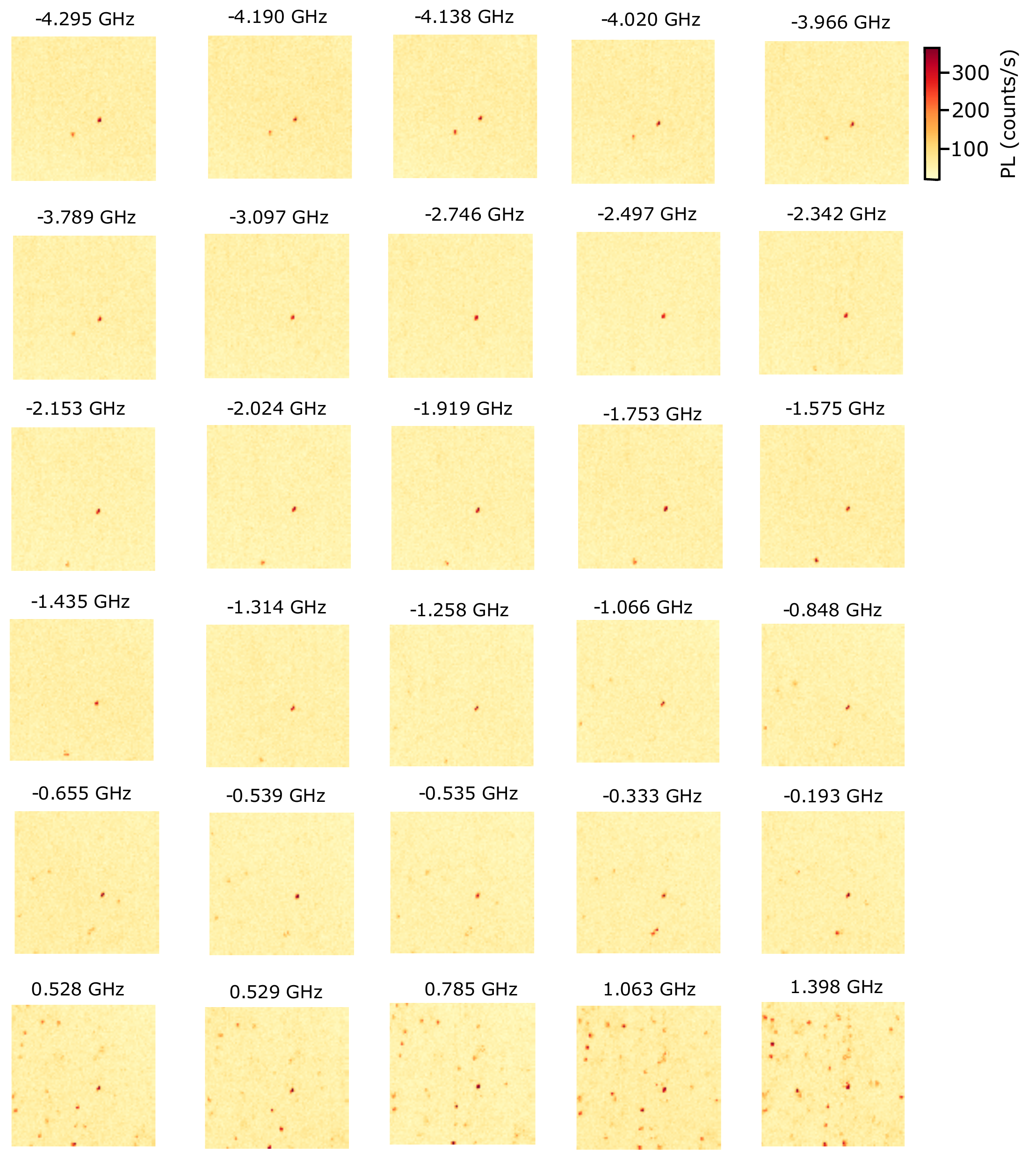}
\caption{\textbf{PLE maps, sample B - natural abundance of isotopes (excitation laser frequency range: \num{-4.3} to \num{+1.4} \unit{\GHz)}.} Photoluminescence maps for different detunings of the excitation laser for the sample with natural abundance of Si and C isotopes. Each map shows a \num{30} \unit{\micro\meter} by \num{30} \unit{\micro\meter} area. A PL spot is visible around the centre of all maps: its PLE spectrum is very broad and not associated to a V centre. For these measurements, we use an integration time of \SI{1}{\second} per step, an excitation power of \SI{14}{\micro\watt} for the repump laser, and \SI{4}{\micro\watt} for the telecom laser.}
\label{fig:map_NA1}
\end{figure*}

\begin{figure*}[!ht]
\includegraphics[width=0.9\textwidth]{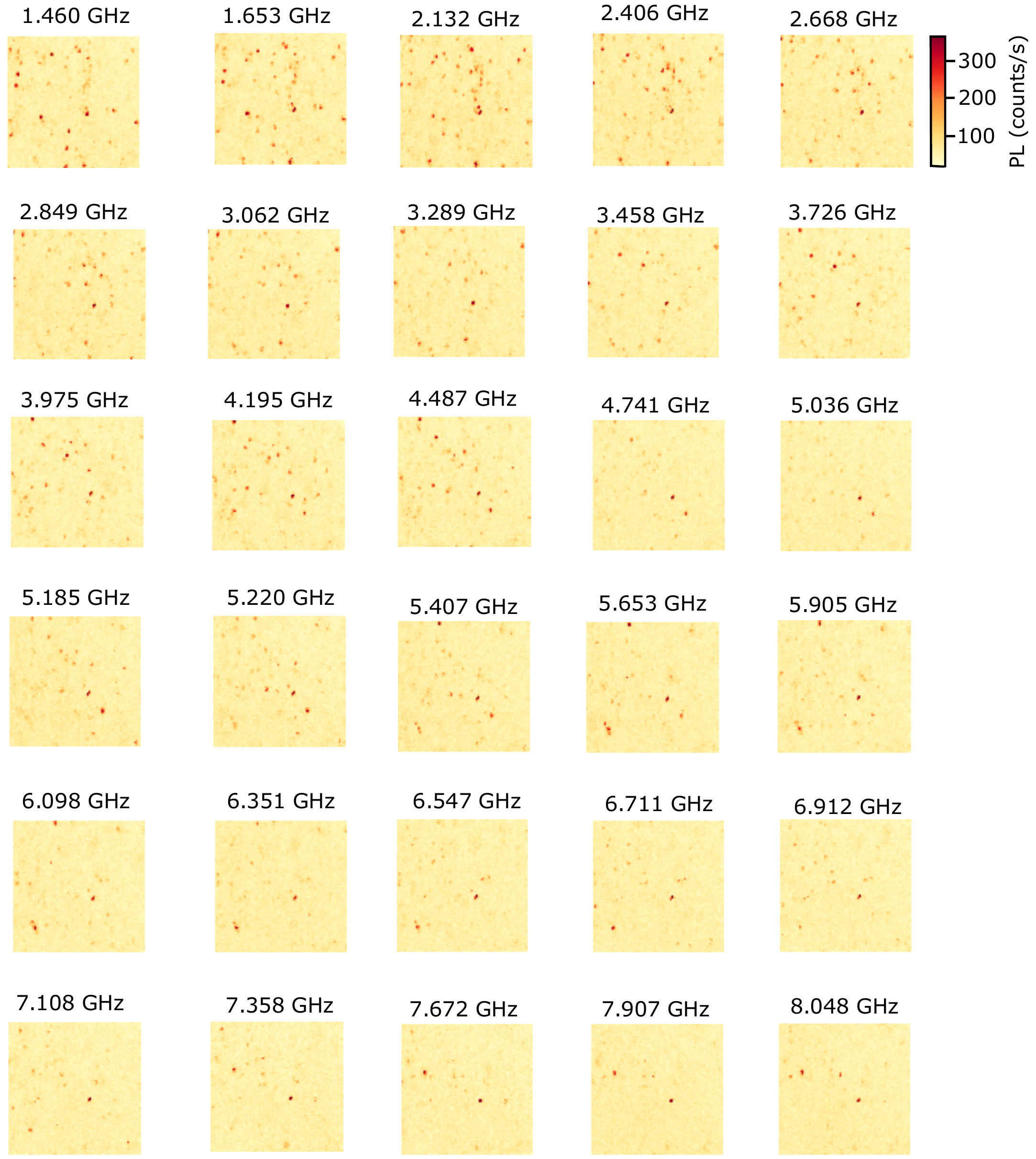}
\caption{\textbf{PLE maps, sample B - natural abundance of isotopes (excitation laser frequency range: \num{1.4} to \num{8} \unit{\GHz}).} Photoluminescence maps for different detunings of the excitation laser for the sample with natural abundance of Si and C isotopes. Each map shows a \num{30} \unit{\micro\meter} by \num{30} \unit{\micro\meter} area. For these measurements, we use an integration time of \SI{1}{\second} per step, an excitation power of \SI{14}{\micro\watt} for the repump laser, and \SI{4}{\micro\watt} for the telecom laser.}
\label{fig:map_NA2}
\end{figure*}

\begin{figure*}[!ht]
\includegraphics[width=0.9\textwidth]{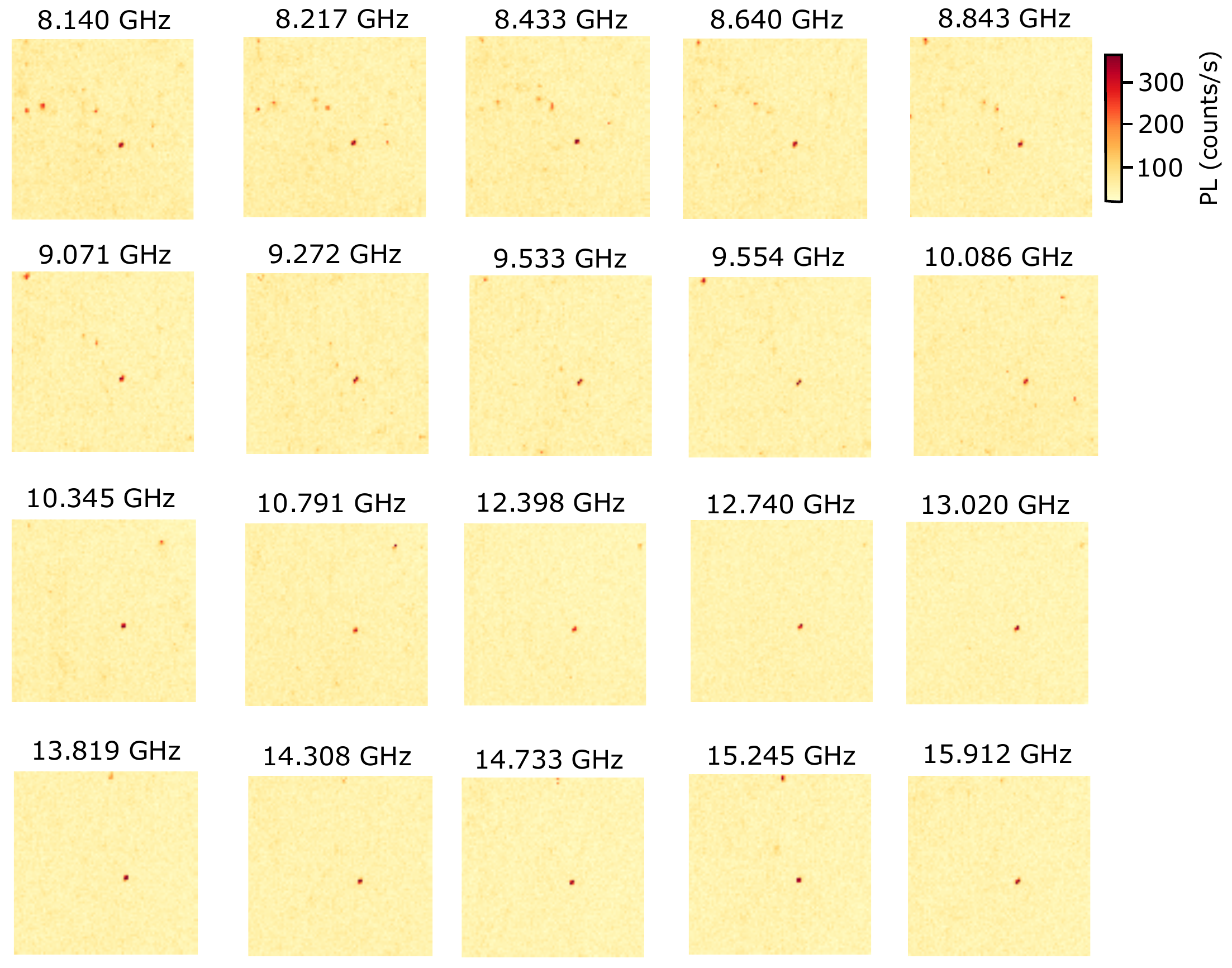}
\caption{\textbf{PLE maps, sample B - natural abundance of isotopes (excitation laser frequency range: 8 to 16 GHz).} Photoluminescence maps for different detunings of the excitation laser for the sample with natural abundance of Si and C isotopes. Each map shows a \num{30} \unit{\micro\meter} by \num{30} \unit{\micro\meter} area. A PL spot is visible around the centre of all maps: its PLE spectrum is very broad and not associated to a V centre. For these measurements, we use an integration time of \SI{1}{\second} per step, an excitation power of \SI{14}{\micro\watt} for the repump laser, and \SI{4}{\micro\watt} for the telecom laser.}
\label{fig:map_NA3}
\end{figure*}

\begin{figure*}[!ht]
\includegraphics[width=1\textwidth]{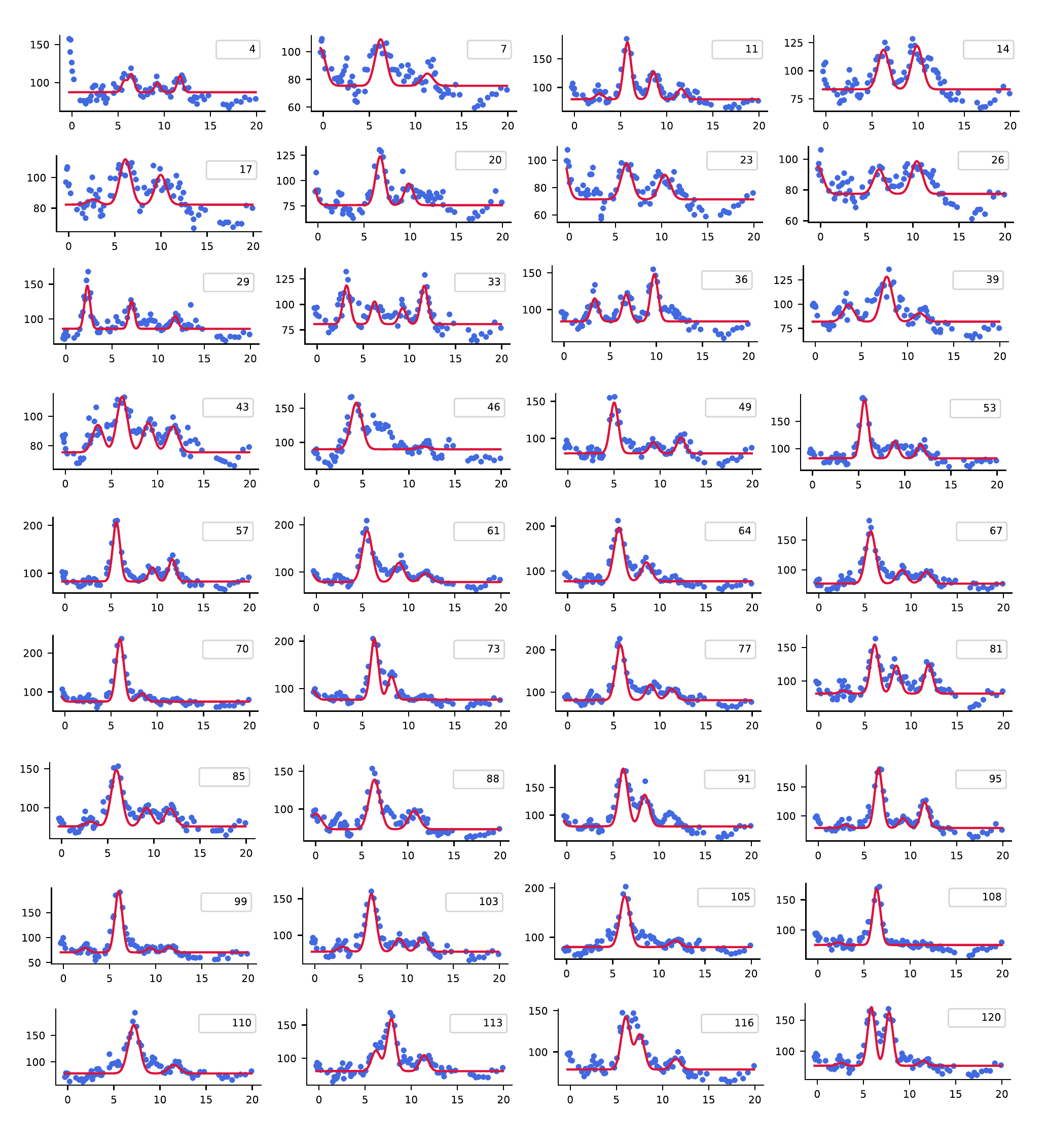}
\caption{\textbf{PLE spectra, sample B: natural abundance of isotopes (part I)}. Fitted spectra from \num{36} PL spots in Supplementary Figure \ref{fig:map_NA2}. For each sub-plot, the x-axis shows the excitation frequency in \unit{\GHz}, and the y-axis the number of detected counts per second.}
\label{fig:spectra_NA1}
\end{figure*}

\begin{figure*}[!ht]
\includegraphics[width=1\textwidth]{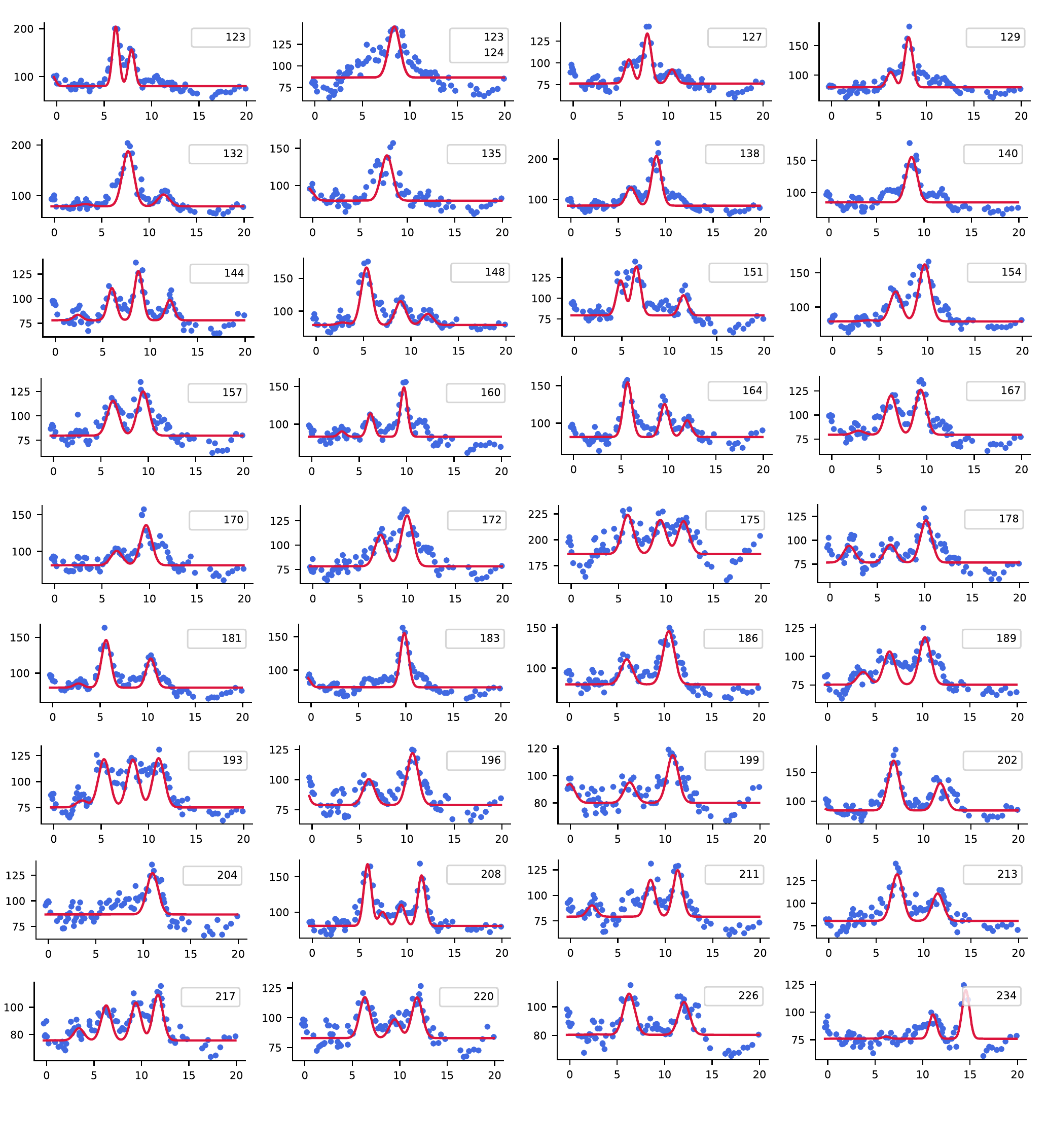}
\caption{\textbf{PLE spectra, sample B: natural abundance of isotopes (part II)}. Fitted spectra from \num{36} PL spots in Supplementary Figure \ref{fig:map_NA2}. PLE spectra for \num{36} more spots are reported in Extended Data 4. For each sub-plot, the x-axis shows the excitation frequency in \unit{\GHz}, and the y-axis the number of detected counts per second.}
\label{fig:spectra_NA2}
\end{figure*}

\begin{figure*}[!ht]
\includegraphics[width=0.95 \textwidth]{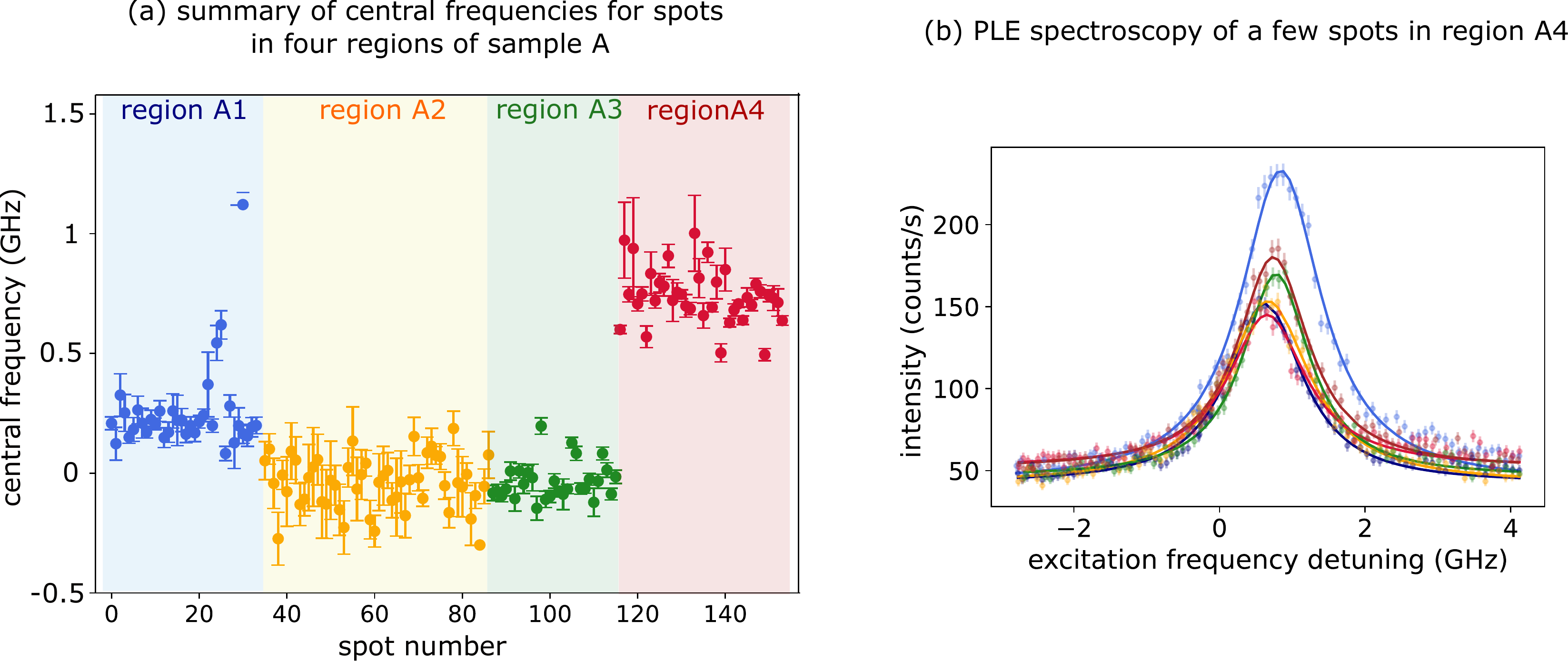}
\caption{\textbf{Inhomogeneous distribution of the vanadium emitters in four regions of the sample.} (a) Summary of central frequencies for \num{150} vanadium centres in four different regions of the sample. In each region, the inhomogeneous distribution features a standard deviation of about $100$ MHz. Data for region A4 were collected after the cryostat was warmed up to room temperature and subsequently cooled down to \num{4} \unit{\K}. The error bars are extracted from the fits of the peaks (i.e., PL spots in the maps). (b) PLE spectroscopy of \num{6} spots in region A4, to confirm the narrow inhomogeneous distribution observed in the sequence of frequency-resolved PLE maps. The error bars correspond to the Poisson noise of the photon counts.}
\label{fig:inhomogeneous_broadening_SI}
\end{figure*}

\clearpage 
\bibliography{Vanadium_Bibliography}

\end{document}